\documentclass[nopacs,twocolumn,preprintnumbers,notitlepage,amsmath,amssymb,superscriptaddress]{revtex4-2}

\usepackage[utf8]{inputenc}
\usepackage{physics}
\usepackage{blindtext}
\usepackage{ dsfont }
\usepackage{tikz}
\usepackage{subfiles}
\usepackage{subcaption}
\usepackage{hyperref}
\usepackage[font=small,labelfont=bf,
   justification=justified,
   format=plain]{caption} 
\graphicspath{ {./Figures/} }
\usepackage{graphicx}
\usepackage{dcolumn}
\usepackage{bm}
\usepackage{amsmath}
\usepackage{amssymb}

\newcommand{\Sim}[1]{\mathrel{\mathop{\kern 0pt\sim}\limits_{#1}}}
\newcommand{\To}[1]{\mathrel{\mathop{\kern 0pt\to}\limits_{#1}}}
\newcommand{\dw}{d_\text{w}}

\makeatletter
\AtBeginDocument{\let\LS@rot\@undefined}
\makeatother
\usepackage[]{pdfpages}

\begin{document}

\title{Aging Record Statistics in Saturating Self-Interacting Random Walks}
\author{J. Br\'emont}
\affiliation{Collège de France, 3 Rue d'Ulm, 75005 Paris, France}
\author{R. Voituriez}
\affiliation{Laboratoire de Physique Th\'eorique de la Mati\`ere Condens\'ee, CNRS/Sorbonne Université, 
 4 Place Jussieu, 75005 Paris, France}
\affiliation{Laboratoire Jean Perrin, CNRS/Sorbonne Université, 
 4 Place Jussieu, 75005 Paris, France}
 \author{O. B\'enichou}
\affiliation{Laboratoire de Physique Th\'eorique de la Mati\`ere Condens\'ee, CNRS/Sorbonne Université, 
 4 Place Jussieu, 75005 Paris, France}
\date{\today}
\begin{abstract}
The record age $\tau_k$, defined as the time between the $k$th and $(k+1)$st record-breaking events, is a central observable of extreme-value statistics. In Markovian processes, the absence of memory makes $\tau_k$ independent of $k$. How memory breaks this invariance and induces aging—a dependence of $\tau_k$ on $k$—remains a fundamental question, closely connected to widely observed aging phenomena in non-Markovian dynamics. In this Letter, we derive the exact asymptotic distribution of $\tau_k$ for saturating self-interacting random walks, a broad class of non-Markovian processes. We uncover two asymptotic regimes, in agreement with recent scaling predictions: at short times ($\tau \ll k^2$), record statistics are governed by the geometry of the explored region, while at long times ($\tau \gg k^2$), memory effects become subdominant and reduce to nontrivial prefactor corrections. Our exact result provides a rare analytic window beyond scaling theory and extends to a framework that fully quantifies aging dynamics in the presence of saturating self-interaction.
\end{abstract}

\maketitle
Record statistics are key observables in the study of extreme events \cite{majumdarExtremeValue}. 
A record occurs at time $t$ when $X_t$ exceeds all previous values. In this context, the 
time intervals $\tau_n$ between the $n$th and $(n+1)$st records, referred to as record ages 
\cite{aliakbariRecordsFractal,benigniHausdorffDimension,ben-naimStatisticsSuperior,
godrecheRecordStatistics,godrecheRecordStatisticsIntegrated2022,
godrecheStatisticsLongest,lacroix-a-chez-toineUniversalSurvival}, are of central importance, 
as they characterize the waiting time until the next record-breaking event. Such statistics 
are relevant for a wide range of phenomena including heatwaves \cite{witzeExtremeHeatwaves}, 
earthquakes \cite{ambraseysValueHistorical}, and record temperatures \cite{coumouGlobalIncrease}.
Record statistics and record ages have been extensively studied \cite{ahsanullahRecordStatistics}, 
starting from the classical case of independent and identically distributed (i.i.d.) time series 
$(X_t)$ \cite{chandlerDistributionFrequency}. More recently, important progress was made 
when the observations were taken as the successive positions of a Markovian random walk (RW), 
$X_{t+1} = X_t + \eta_t$, where the increments $(\eta_t)_{t \ge 0}$ are i.i.d.\ and symmetric 
\cite{godrecheRecordStatistics}. In this case, while the positions $(X_t)$ exhibit long-range temporal 
correlations, independence of the increments implies that the process has no memory 
of its past trajectory. This renewal property leads to translation invariance, implying that 
the record age $\tau_k$ is independent of the record index $k$.
As a result, the distribution of record ages follows a universal algebraic decay
\begin{equation}
\mathbb{P}(\tau_k \ge \tau) \propto \tau^{-\theta},
\end{equation}
where $\theta$ is the persistence exponent \cite{Bray:2013,majumdar}. For symmetric random walks 
with independent increments, the Sparre--Andersen theorem \cite{andersenNumberPositive} gives 
$\theta = 1/2$. Therefore, despite the correlations in the positions, the absence of memory in 
the increments prevents any aging of records in the Markovian case, and the statistics of record 
ages do not depend on the record index $k$.

This picture changes fundamentally in many real-world systems where non-Markovian effects are ubiquitous. 
In contexts ranging from foraging behavior to financial markets and intracellular transport, 
the dynamics often retain memory of past evolution, so that future behavior depends not only on the 
current state but on the entire history of the system \cite{tarasovFractionalDynamics,cellattract,
saxtonAnomalousDiffusion,barkaiAreaBessel,naturealex,trogerSizedependentSelfavoidance,boyerRandomWalks,ants}. 
In such situations, reaching a given record level $k$ carries information about the past trajectory, 
which in turn affects the subsequent evolution.
An important first step towards understanding record ages in presence of memory was made in \cite{records}, which developed a scaling theory for record ages in generic non-Markovian, symmetric, scale-invariant random walks $(X_t)$. Define the walk dimension $\dw$ by the scaling behavior $X_t \propto t^{1/\dw}$; for example, $\dw=2$ for Brownian motion \cite{hughes}. The survival probability $S_k(\tau)\equiv \mathbb{P}(\tau_k \ge \tau)$ was predicted to obey, in the joint limit $\tau,k\to\infty$ with $x=k^{-\dw}\tau$ fixed, the scaling form
\begin{equation}
   \label{scaling-leo}
   S_k(\tau) \sim k^{-1}\,\psi(x),
\end{equation}
where $\psi$ is a scaling function that depends on the underlying dynamics. Remarkably, $\psi$ was further predicted to display two universal regimes \cite{records}:
\begin{equation}
   \label{regimes-leo}
   \psi(x) \propto 
   \begin{cases}
      x^{-1/\dw}, & x \ll 1,\\
      x^{-\theta}, & x \gg 1.
   \end{cases}
\end{equation}
For a simple Markovian RW such as Brownian motion, $\psi(x)=\sqrt{2/(\pi x)}$ \cite{hughes}, and the two regimes in \eqref{regimes-leo} coincide. This reflects the absence of aging: $\tau_k$ is independent of $k$. The distinction between these two regimes is therefore a consequence of long-range memory effects. However, no microscopic derivation of the record age distribution $S_k(\tau)$—nor even of the emergence of its two regimes—has so far been obtained for RWs with long-range memory.
By contrast, record statistics are well understood for renewal \cite{godrecheStatisticsLongest,godrecheRecordStatisticsIntegrated2022,majumdarUniversalFirstpassageProperties2010}, short-range correlated \cite{lacroix-a-chez-toineUniversalSurvival}, and time-dependent Markov \cite{records} RWs.

More broadly, the determination of $\psi$ and the emergence of its two regimes are closely linked to aging in non-Markovian dynamics, which has recently attracted significant interest \cite{schulzAgingRenewal,bremontAgingDynamics,Barbier:2022,foxAgingPower, bremontFlipsRevealb,barkaiAreaBessel,boyerRandomWalks}. Self-interacting RWs (SIRWs), which modify and subsequently interact with their previously visited environment, provide a natural setting where such non-Markovian effects emerge. In these systems, moving agents leave persistent footprints that influence subsequent motion, which may be of chemical \cite{naturealex,cellattract,trogerSizedependentSelfavoidance,hokmabadChemotacticSelfcaging,chenEvolvingMotility,kranzEffectiveDynamics}, mechanical \cite{adar}, or informational origin (such as in nonreversible Monte Carlo algorithms \cite{massoulieVelocityTrappinga,maggs1}).
While standard, non-aged observables such as propagators and FPTs are now increasingly well understood \cite{bremontExactPropagators,bremontPersistenceExponentsa,dumazMarginalDensitiesTrue2012,Barbier:2020,massoulieVelocityTrappinga,romano2026anomalous}, numerical and experimental studies reveal distinctive aging behavior that still lacks a quantitative description \cite{Barbier:2022,trogerSizedependentSelfavoidance,cellattract,naturealex}, such as the emergence of new scaling exponents.
An exact computation of $\psi$ for a genuinely non-Markovian RW such as a SIRW would serve three key purposes. First, it would place the scaling framework~\cite{records} on firm microscopic grounds and clarify its underlying assumptions and limits \footnote{In particular, it would make explicit the effective independence of record ages when considering extreme statistics.}. Second, it would provide access to the prefactors in each regime, which are process-dependent and cannot be obtained from scaling arguments alone, making them relevant for comparison with experiments. Finally, it would describe the full crossover between the two regimes, providing physical insight into how aging emerges from self-interaction, from an early-time regime dominated by memory of the past trajectory to a regime where its influence becomes negligible. 
In this Letter, we achieve this goal by introducing a general framework for aging dynamics in saturating SIRWs (exemplified by the SATW model defined below), yielding in particular the exact scaling function $\psi$ governing their record ages.

\begin{figure}
    \centering
      \includegraphics[width=\columnwidth]{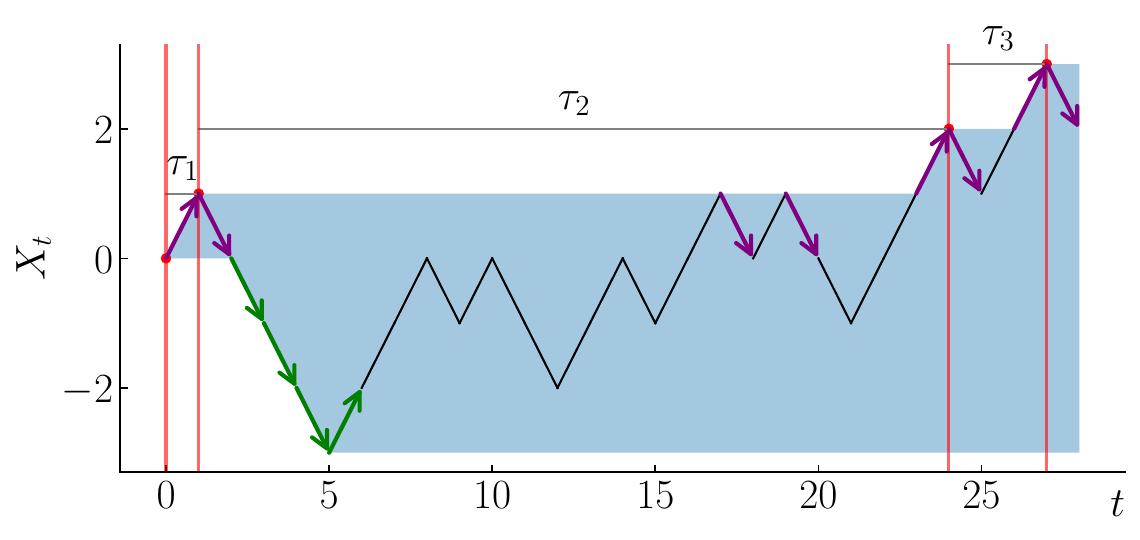}
    \caption{Trajectory and record ages $\tau_k$ of a SATW. The growing region visited by the walker is represented in blue. Steps taken starting from a local maximum (resp. minimum) are represented by purple (resp. green) arrows, while black steps in the bulk of the visited region have probability $1/2$, as the walker behaves as a simple RW in this region. Self-interaction is present at the boundaries, where each upward (resp. downward) purple arrow has probability $1/(1+\alpha)$ (resp. $\alpha/(1+\alpha)$), while each upward (resp. downward) green arrow has probability $\beta/(1+\beta)$ (resp. $1/(1+\beta)$).}
    \label{fig:satw-dessin}
\end{figure}
The SATW is a non-Markovian nearest-neighbor RW $(X_t)$ on $\mathbb{Z}$, starting at $X_0=0$, in which transition probabilities depend on whether neighboring sites have been previously visited. If the walker at site $x$ has not visited $x+1$ (resp.\ $x-1$), it jumps to this site with probability $1/(1+\beta)$ (resp.\ $1/(1+\alpha)$), and to the visited site with complementary probability. If both neighbors have been visited, the walk jumps left or right with probability $1/2$ (see Fig.~\ref{fig:satw-dessin}). The case $\alpha=\beta=1$ reduces to the simple RW, while $\alpha>1$ (resp.\ $\beta>1$) induces attraction toward previously visited sites on the left (resp.\ right), and $\alpha<1$ (resp.\ $\beta<1$) corresponds to repulsion. We use asymmetric weights $\alpha \neq \beta$ to keep the formulation general and 
to disentangle the respective roles of the left and right boundaries in the expression of our results.

More generally, we consider edge-reinforced SIRWs with transition probabilities $\rho_{x,x\pm1}\propto w_{\mathrm{sign}(x)}(n_{x,x\pm1})$, where $n_{x,x\pm1}$ is the number of crossings of edge $\{x,x\pm1\}$. When self-interaction saturates, i.e. $w_\pm(n)\sim l_\pm + \mathcal{O}(n^{-1})$ with $l_\pm >0$, the dynamics become equivalent to a SATW with effective parameters $\alpha,\beta$ given by \cite{Toth:1996}
\[ \alpha = w_-(0)^{-1} + \sum_{j\ge 1} \left[ w_-(2j)^{-1}-w_-(2j-1)^{-1} \right], \]
and analogously $\beta$ defined from $w_+$. Physically, the walk is effectively Markovian in the bulk of the visited interval, where edges are frequently crossed so that $w_\pm(n)$ is independent of $n$, while memory persists near the boundaries, where edges are visited only $\mathcal{O}(1)$ times \footnote{Similar conclusions hold for site-reinforced SIRWs \cite{bremontExactPropagators}.}. Our results therefore extend to the broad universality class of saturating SIRWs, recently observed experimentally~\cite{naturealex,cellattract,trogerSizedependentSelfavoidance}.


\begin{figure}
    \centering
    \begin{subfigure}[t]{0.49\columnwidth}
        \includegraphics[width=1\textwidth]{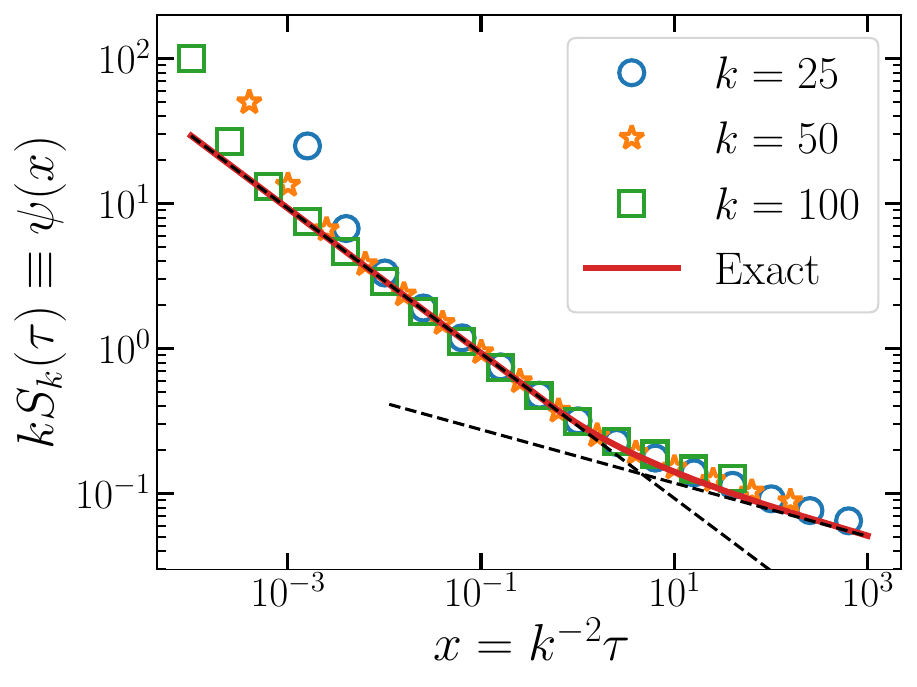}
        \label{fig:records-phiem1}
    \end{subfigure}
   \hfill
    \begin{subfigure}[t]{.49\columnwidth}
        \includegraphics[width=1\textwidth]{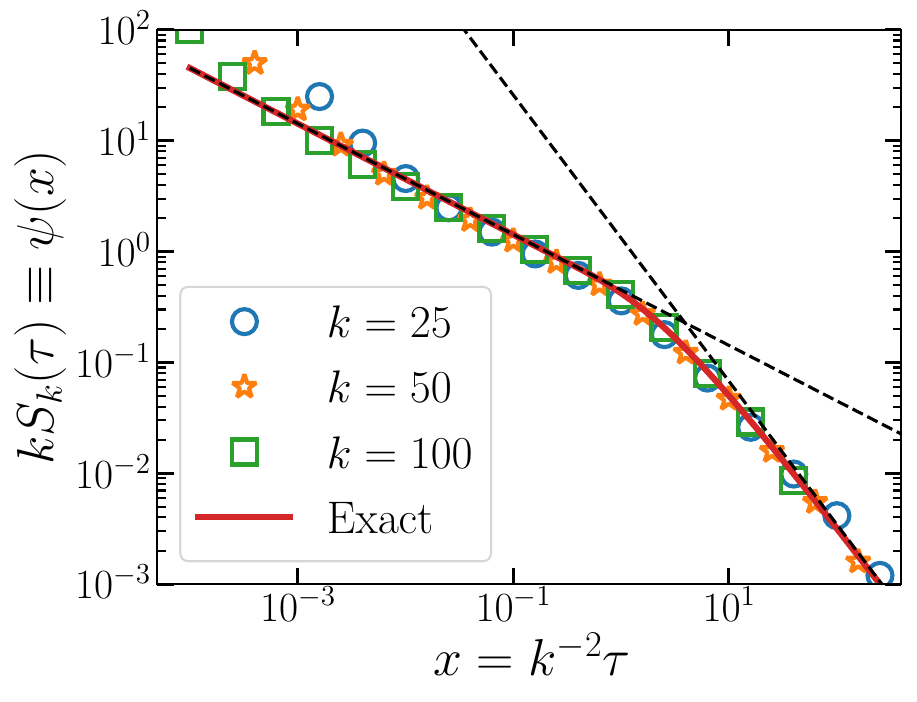}
        \label{fig:records-asym}
    \end{subfigure}
   \vspace{-3ex}
    \caption{Distribution $S_k(\tau)$ of the $k$th record age $\tau_k$. 
\textbf{Left:} self-repelling, symmetric SATW with $\alpha=\beta=e>1$. 
\textbf{Right:} mixed asymmetric SIRW with a side-dependent, saturating weight function 
$w_{\pm}(n)=\phi_{\pm}^{-1}+(1-\phi_{\pm}^{-1})\!\left(2[1+e^{-n}]^{-1}-1\right)$,
with $\phi_+=1/6$ and $\phi_-=3$. 
The red curve shows the exact theoretical prediction, Eq.~\eqref{psi}, while symbols denote numerical simulations for different values of $k$. Dashed black lines indicate the exact small-$x$ and large-$x$ asymptotics, Eq.~\eqref{asymptotics-psi}. 
The two scaling regimes, clearly visible and separated by only a narrow crossover region (less than one decade in the rescaled variable $x$), are characteristic signatures of non-Markovian dynamics. 
The excellent agreement in the right panel demonstrates that our main result Eq.~\eqref{psi} holds beyond the minimal SATW model, confirming robustness across the entire university class of saturating SIRWs. 
}
    \label{fig:records}
\end{figure}

The main result of this Letter is the exact asymptotic distribution of record ages for saturating SIRWs, via the following explicit expression for the scaling function $\psi$ defined by Eq.~\eqref{scaling-leo}:
\begin{align}
   \label{psi}
    \psi\left(x\right) =
\beta\,\sqrt{\frac{2}{\pi x}} \frac{\int_0^1 u^{\alpha-1}(1-u)^{\beta-1}\left(2\, \Xi_{\alpha}(x u^2) -1 \right) \,du}{B(\alpha,\beta)},
\end{align}
where $B(\alpha,\beta)=\int_0^1 u^{\alpha-1} (1-u)^{\beta-1} \, du$ is the Beta function, and we introduced the auxiliary function
\begin{equation}
   \label{def-Xi}
   \Xi_{\alpha}(u) = \sum_{n=0}^\infty \frac{\left(\frac{1-\alpha}{2} \right)_n}{\left(\frac{1+\alpha}{2} \right)_n} \, \exp(-\frac{2n^2}{u}),
\end{equation}
where $(x)_n = x(x+1)\dots(x+n-1)$ is the Pochhammer symbol. 

Our main result Eq.~\eqref{psi} is easily evaluated numerically: see Fig.~\ref{fig:records} for a confirmation against numerical simulations. 
Its asymptotic behavior, derived in the SM, becomes physically interpretable when expressed in terms of the record-age distribution:
\begin{equation}
   \label{asymptotics-psi}
   S_k(\tau) \sim 
   \begin{cases}
      \displaystyle \beta \sqrt{\frac{2}{\pi \tau}}, 
      & 1 \ll \tau \ll k^2, \\[6pt]
      \displaystyle 
      k^{\alpha-1} 
      \frac{\Gamma\!\left( \frac{1 + \alpha}{2} \right)}
      {\sqrt{\pi} \, B(\alpha, \beta)} 
      \left( \frac{2}{\tau} \right)^{\alpha/2}, 
      & \tau \gg k^2,
   \end{cases}
\end{equation}
where $\Gamma$ is the Gamma function.

This central result Eq.~\eqref{psi} calls for several comments.
(i) To our knowledge, this constitutes the first derivation of record-age statistics for SIRWs, and more generally for any non-Markovian RW with long-ranged memory \footnote{In the mathematical literature, we are aware of a single related study \cite{serletExplicitLaws} which, while seemingly addressing record statistics of the SATW, actually concerns the number of distinct visited sites. It does not contain or imply our main result Eq.~\eqref{psi}.}. It is in excellent agreement with numerical simulations, both for the minimal SATW model and for more complex saturating SIRWs in the SATW universality class (see Fig.~\ref{fig:records}). For SATW, one has $\dw=2$ \cite{Toth:1996} and $\theta=\alpha/2$ \cite{bremontPersistenceExponentsa,Barbier:2020}: the two asymptotic regimes in \eqref{asymptotics-psi} thus provide a nontrivial exact confirmation of the scaling prediction Eq.~\eqref{regimes-leo} from microscopic non-Markovian dynamics. 

(ii) The $\alpha=1$ case leads to a drastic simplification: since $\Xi_{1}(u) \equiv 1$, Eq.~\eqref{psi} becomes $\psi(x)=\beta\sqrt{\tfrac{2}{\pi x}}$. This simplification can be traced back to a simple renewal picture, which provides important physical intuition. For $\alpha=1$, the walker behaves as a simple RW on the half-space $x<k$. Starting from $k$ at time $T$, the record is broken at a time larger than $T+\tau$ with the walker visiting $k$ exactly $n$ times with probability
\[
\left(\frac{\beta}{1+\beta}\right)^n \frac{1}{1+\beta}
\, \mathbb{P}(t_1+\dots+t_n\geq \tau),
\]
where $t_i$ are independent return times of $1d$ Brownian motion, with common density $\rho(t)$.  
Using the Laplace transform $t\to s$ defined for a function $f(t)$ by $\hat{f}(s) = \int_0^\infty e^{-s t} f(t) dt$, it is well-known \cite{hughes} that $\hat{\rho}(s)=e^{-\sqrt{2s}}$. We thus obtain
\begin{equation*}
   s\hat{S}_{k}(s)
   =1-
   \frac{1}{1+\beta}
   \sum_{n=0}^\infty
   \left(\frac{\beta e^{-\sqrt{2s}}}{1+\beta}\right)^n
   \Sim{s\to0}
   1-e^{-\beta \sqrt{2s}}.
\end{equation*}
Laplace inversion directly yields $S_k(\tau)\Sim{\tau \gg 1}\beta\,\sqrt{\tfrac{2}{\pi\tau}}$. For $\alpha\neq1$, this renewal picture breaks down, and $\Xi_\alpha \neq 1$, reflecting the influence of the past. It is worth mentioning that when $\alpha = 2k+1$ is an odd integer, the series \eqref{def-Xi} defining $\Xi_\alpha$ becomes a finite sum, yielding simple expressions for $\psi$ (see SM) : a similar behavior was observed for the propagator of SATW in \cite{bremontExactPropagators}.

(iii) The scaling function $\psi(x)$ admits a simple physical interpretation when written as an 
average over a $\mathrm{Beta}(\alpha,\beta)$ distributed random variable $U$:
\begin{equation}
   \label{psi-mean}
   \psi(x) = \biggl \langle \beta \sqrt{\frac{2}{\pi x}} \left(2\, \Xi_{\alpha}(x U^2) -1 \right) \biggr \rangle.
\end{equation}
The variable $U$ corresponds to the aspect ratio $U = k/(k+m)$ of the visited interval 
$[-m,k]$ when site $k$ is first reached \cite{bremontPersistenceExponentsa} (see also 
Eq.~\eqref{dist-m1} for the distribution of $m = k(1-U)/U$). Up to a factor $k$ 
\eqref{scaling-leo}, the term inside the brackets in Eq.~\eqref{psi-mean} can be interpreted 
as the probability that, starting from $k$ with minimum $-m$, record $k$ has not yet been 
broken at time $\tau$. Interestingly, its dependence on the right boundary enters only 
through a multiplicative factor $\beta$, whereas nontrivial memory effects arise solely 
from the left boundary through $\Xi_\alpha$, which is independent of $\beta$. The function $\psi$ is finally obtained by averaging over the single hidden degree of freedom $m$.
Since $x=\tau/k^2$, the argument of $\Xi_\alpha$ becomes $xU^2=\tau/(k+m)^2$, which introduces 
the diffusive timescale $T=(k+m)^2$ corresponding to the time needed for the walker starting 
from $k$ to feel the lower boundary $-m$. For $\tau \ll T$, the walker behaves as if the region 
below $k$ were effectively semi-infinite and one recovers the $\alpha=1$ regime. For 
$\tau \gtrsim T$, the walker feels the boundary and the full function $\Xi_\alpha$ is involved. 
Aging in record statistics of SATW therefore originates from memory effects localized at the boundaries 
of the explored region.

(iv) Our approach naturally extends to aged FPTs. Rather than the classical FPT $T_x$ to a fixed target, one may consider its aged counterpart: given that the walker has already reached a site $k$, how long does it take to reach a site $x>k$? This aged FPT, denoted $T_k^x$, can be expressed as
\begin{equation}
\label{aged-fpt}
T_k^x=\sum_{j=k}^{x-1}\tau_j,
\end{equation}
showing that record ages $\tau_j$ constitute the elementary building blocks of first-passage events. Note that aged FPTs reduce to classical FPTs for Markovian RWs. In turn, classical (non-aged) FPTs in non-Markovian RWs such as SATW have been studied previously~\cite{bremontPersistenceExponentsa,Barbier:2020,carmonaBetaVariables,coghiAcceleratedFirstPassagea,wiese,guerinMeanFirstpassage}, but their aged counterparts remain largely unexplored. As detailed below, our method provides access to the full asymptotic distribution of $T_k^x$. Here, we focus on the record age limit $x=k+1$ which already captures the essential physics and admits simpler expressions.

\emph{Derivation of the scaling function $\psi$ for SATW.} We now derive our main result~\eqref{psi}, namely the statistics of the record ages $\tau_k$ for the SATW. 
Consider a trajectory that reaches site $k>0$ for the first time at time $T_k$. We want to know at which time $T_k +\tau_k$ it will first hit $k+1$.
The main difficulty stems from the non-Markovian nature of the SATW: the evolution at times $t>T_k$ depends on the entire past trajectory. 
Because the defining feature of SATW is that the walker behaves as a simple RW in the bulk of its visited territory, this dependence can be encoded in a single hidden variable: the position $-m$ of the leftmost site visited at time $T_k$. Our derivation can thus be decomposed in three key steps. (S1) compute the distribution of the left boundary $-m$, (S2) quantify how the past trajectory, i.e. a given value of $-m$, impacts the record age $\tau_k$, and (S3) average the result of (S2) with respect to $-m$. 

(S1) For completeness, we briefly recall the derivation of the statistics of $-m$ following ~\cite{bremontPersistenceExponentsa}. 
The central object is the local time process $(L_{T_k}(x))_{x\in\mathbb Z}$, where $L_{T_k}(x)$ denotes the total number of visits to site $x$ before time $T_k$. 
In particular, $L_{T_k}(k)=1$, and we have the identity
\begin{equation}
\label{area-loctime}
\sum_{x\in\mathbb Z} L_{T_k}(x)=T_k,
\end{equation}
which simply expresses that each time step contributes one unit of local time somewhere. 
We next rely on two key properties of the local time process.
(i) Its support coincides with the set of visited sites at time $T_k$. 
Consequently, $-m$ is precisely the leftmost site $x$ such that $L_{T_k}(x)>0$, and $-m-1$ can be interpreted as a first-passage to $0$ of the local time process. 
(ii) In the scaling limit, it admits an explicit, so-called \emph{Ray--Knight} description, a result first obtained by T\'oth~\cite{Toth:1996}. 
More precisely, introducing the rescaled coordinate $x=Ax'$ and taking the joint limit $A\to\infty, x'\to 0$ with $x$ fixed, one obtains
\begin{equation}
\label{rayknight}
\frac{L_{Ax'}(T_{Ak'})}{A} \To{A\to\infty}
\begin{cases}
Y_{2\beta}(k'-x'), & 0<x'<k',\\[4pt]
\tilde Y_{2-2\alpha}(-x'), & x'<0,
\end{cases}
\end{equation}
where $Y_\delta$ denotes a squared Bessel process of dimension $\delta$ (or BESQ$_\delta$), defined as follows. When $\delta \in \mathbb{N}$, the BESQ$_\delta$, denoted $Y_\delta$, is simply the squared norm $||\vec{B}_t||^2$ of a
$\delta$-dimensional Brownian motion $\vec{B}_t$. For noninteger values
of $\delta$, it is the solution $Y_\delta$ of the stochastic differential
equation 
\begin{equation}
   dY_\delta(t) = \delta dt + 2 \sqrt{|Y_\delta(t)|} \,dW_t, 
\end{equation}
where $dW_t$ is uncorrelated Gaussian white noise with variance $1$ ~\cite{going-jaeschkeSurveyGeneralizations}. The processes $Y_{2\beta}$ and $\tilde Y_{2-2\alpha}$ in Eq.~\eqref{rayknight} satisfy the boundary conditions $Y_{2\beta}(0)=0$ and
$\tilde Y_{2-2\alpha}(0)=Y_{2\beta}(k)$, which reflect continuity of the local time process. Importantly, they are independent once conditioned on this common value. Finally, the tilde in $\tilde Y_{2-2\alpha}$ means that this process is absorbed upon hitting zero. For notational simplicity we henceforth drop primes ($x'\to x$), keeping in mind that~\eqref{rayknight} holds in the scaling limit.

We first remind that, using the decomposition \eqref{rayknight}, the distribution $q_+(m,k)$ of the minimum $-m$ at time $T_k$ or, equivalently, of the FPT $-m$ to zero of the process $\tilde Y_{2-2\alpha}$ was shown to be \cite{bremontPersistenceExponentsa}
\begin{equation}
   \label{dist-m1}
q_+(m,k)=\frac{1}{B(\alpha,\beta)}\,
\frac{k^\alpha m^{\beta-1}}{(k+m)^{\alpha+\beta}}.
\end{equation}

(S2) We now ask: how does the memory of the past, fully encoded by the knowledge of the minimum $-m$, influence the time $\tau_k$ needed for the walker starting from $k$ to reach $k+1$ ? To answer this question, we are naturally led to study the more general problem of the aged dynamics of a SIRW which has already explored a finite interval. 
Beyond the context of record ages, understanding this aged regime is important for describing aging phenomena observed, for example, in cell motility and self-propelling microdroplet experiments~\cite{naturealex,cellattract,chenEvolvingMotility,hokmabadChemotacticSelfcaging}, as well as in numerical simulations of SATW~\cite{Barbier:2022,fosterReinforcedWalks}.
Hence, the more general question we address in step (S2) is: 
given that at time $T_{k_1}$ the walker has explored the interval $[-m_1,k_1]$, what is the aged FPT $T_{k_2}$ required to reach the new level $k_2>k_1$? 
Note that the particular case $k_2=k_1+1$ reduces to the record age $\tau_{k_1}$.

Our starting point is a direct two-time extension of the identity~\eqref{area-loctime},
\begin{equation}
   \label{area-aged}
T_{k_2}-T_{k_1}
=\sum_{x\in\mathbb Z}\!\left[L_{T_{k_2}}(x)-L_{T_{k_1}}(x)\right],
\end{equation}
which expresses the elapsed time between the two hitting events as the total increase of local time over all sites.
Therefore, instead of describing the full local time process at time $T_{k_2}$, the aged FPT $T_{k_2}-T_{k_1}$ is obtained by keeping track of the local time increment between two first-passage events:
\begin{equation}
I_x(k_2\,|\,k_1,m_1)
\equiv
L_{T_{k_2}}(x)-L_{T_{k_1}}(x),
\end{equation}
\begin{figure}
    \centering
      \includegraphics[width=\columnwidth]{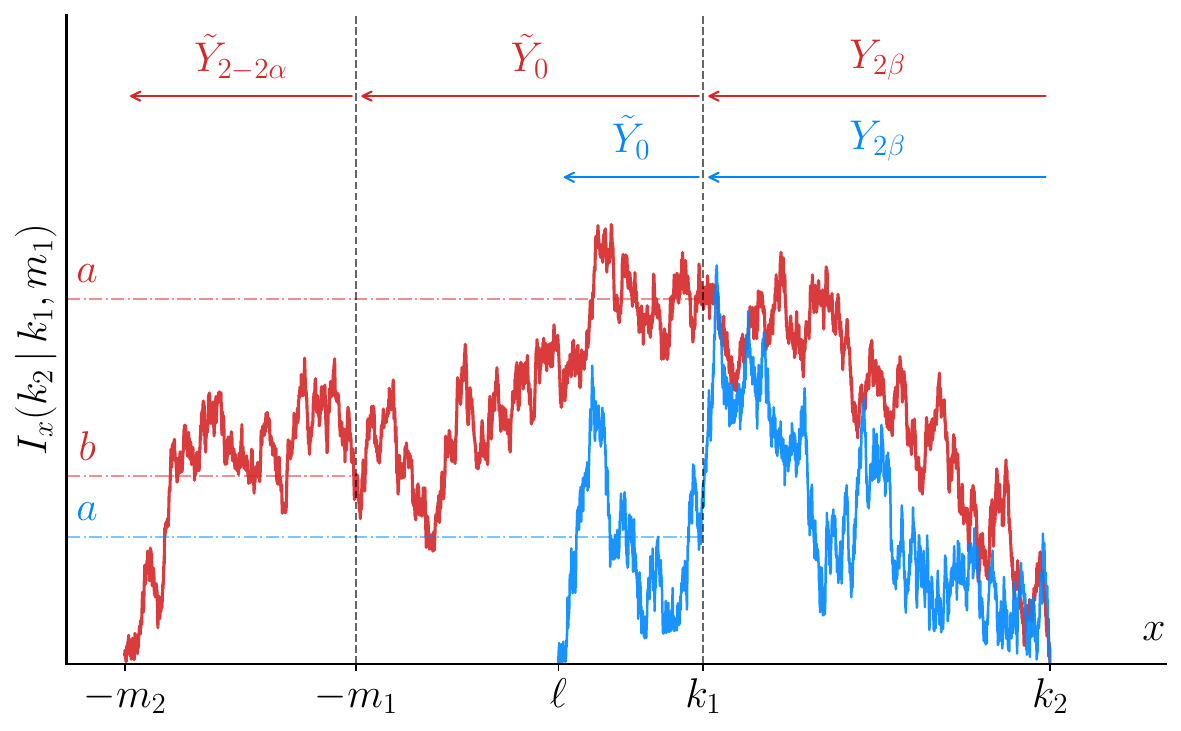}
    \caption{Illustration of the aged Ray--Knight theorem Eq.~\eqref{aged-rk} obtained in \cite{chaumontUpperLower}, used here to quantify aging dynamics in the SATW. The variables $a,l,b\dots$ shown here are the same as in Eq.~\eqref{full-p-A-laplace}. This figure shows two possible local time increments (red and blue) $I_x(k_2 \, | \, k_1, m_1)$ when reaching $k_2$, after discovering an interval $[-m_1, k_1]$. The red increment shows the case where the walker goes beyond its past minimum $-m_1$ when reaching $k_2$. In terms of local time increments, $I_x(k_2 \, | \, k_1, m_1)$ has not hit zero on $[-m_1, k_1]$. The blue increment illustrates the other possible scenario where the walker has not gone beyond $-m_1$ when reaching $k_2$, and the local time increment is strictly zero beyond some site $-m_1<\ell<k_1$. In terms of local time increments, both $\ell$ and $-m_2$ are FPTs to zero of $I_x(k_2 \, | \, k_1, m_1)$.}
    \label{fig:aged-rk}
\end{figure}
We now need an analogue of the Ray--Knight decomposition~\eqref{rayknight} for the increment process $(I_x(k_2\,|\,k_1,m_1))_{x \in \mathbb{Z}}$. While a priori complicated for a general SIRW, this task proves tractable for the SATW, and more generally for the universality class of saturating SIRWs. Again, the key point is that the walker behaves as a simple RW in the bulk of the visited territory: the statistics of the increment process $(I_x(k_2\,|\,k_1,m_1))_{x\in\mathbb Z}$ are thus entirely determined by the support of the previously visited region, encoded by the single parameter $m_1$. In the scaling limit (primed variables omitted for simplicity), it has been shown in \cite{chaumontUpperLower} that
\begin{equation}
   \label{aged-rk}
I_{x}(k_2\,|\,k_1,m_1) \sim
\begin{cases}
Y_{2\beta}(k_2-x), & k_1 < x < k_2,\\
\tilde{Y}_{0}(k_1-x), & -m_1 < x< k_1,\\
\tilde{Y}_{2-2\alpha}(-x), & x<-m_1,
\end{cases}
\end{equation}
where, as in Eq.~\eqref{rayknight}, the three BESQ become independent once conditioned on their matching boundary values (see Fig.~\ref{fig:aged-rk}), 
\begin{equation}
   \label{bc-aged-rk}
   Y_{2\beta}(k_2-k_1)=\tilde Y_0(0), \quad \tilde Y_0(k_1)=\tilde Y_{2-2\alpha}(0).
\end{equation}
This important aged extension of the classical Ray–Knight decomposition \cite{Toth:1996} for SATW has been used in the mathematical literature to prove scaling limits \cite{kosyginaConvergenceRandom} or to obtain bounds on the SATW trajectory \cite{chaumontUpperLower}. Here, we show that Eq.~\eqref{aged-rk} actually provides a practical framework to fully quantify aged observables such as record statistics and aged FPTs, using the known properties of BESQ \footnote{Note that \eqref{aged-rk} is easily iterated by considering more than two successive visited territories, fully characterizing aging in SATW.}. 


(S3) Finally, we illustrate the use of \eqref{aged-rk} in practice by computing the distribution of the aged FPT $T_{k_2}-T_{k_1}$ (recall \eqref{aged-fpt} that the record age $\tau_k$ is obtained by setting $k_2=k_1+1=k+1$). To do so, we average the distribution $P_{\mathrm{incr}}(\mathcal{A})$ of the area $\mathcal{A}=\sum_{x \in \mathbb{Z}} I_x(k_2\,|\,k_1,m_1)$ under the increment process over the past history, embodied by the single variable $-m_1$ \eqref{dist-m1}:
\begin{equation}
   \label{sktau-final}
   \mathbb{P}(T_{k_2}-T_{k_1}\geq \tau)= \sum_{m_1=0}^\infty q_+(m_1,k) \sum_{t=\tau}^\infty P_{\mathrm{incr}}\left(\mathcal{A}=t\right).
\end{equation} 
Thanks to \eqref{aged-rk}, we can write the total area as
\begin{equation}
   \label{tot-area-sum}
   \mathcal{A}=\mathcal{A}_{2\beta}+\tilde{\mathcal{A}}_{0}+\tilde{\mathcal{A}}_{2-2\alpha},
\end{equation}
i.e. as the sum of the respective areas under each of the three BESQ appearing in Eq.~\eqref{aged-rk} (see also Fig.~\ref{fig:aged-rk}). Recall that, conditionnally on the matching boundary conditions \eqref{bc-aged-rk}, these three processes are independent. We thus need to keep track of (i) their respective areas and (ii) their endpoints; that is, one needs the densities 
\begin{align}
   \label{f-p-besq}
   P_\delta^x(b, \mathcal{A}_{\delta} \,|\, a) &\equiv P\left(Y_\delta(0) = a,\, Y_\delta(x) = b, \int_0^x Y_\delta(u) \, du = \mathcal{A}_\delta\right) \nonumber \\ 
   F_\delta^x(\tilde{\mathcal{A}}_{\delta} \,|\, a) &\equiv P\left(\tilde{Y}_\delta(0) = a,\, T_0 = x, \int_0^x \tilde{Y}_\delta(u)  \, du = \tilde{\mathcal{A}}_\delta\right),
\end{align}
where $T_0$ is the FPT of $\tilde{Y}_\delta$ to $0$. Thanks to the convolution structure enabled by the
decomposition of the total area Eq.~\eqref{tot-area-sum},
it is natural to work in Laplace space $\mathcal A\to s$,
where convolutions factorize. We denote transforms of
functions $f(\mathcal A)$ by $\hat f(s)$. While $\hat{P}_\delta^x$ can be found in standard references (see e.g. \cite{borodinHandbookBrownian}), it is not the case for its counterpart $\hat{F}_\delta^x$: we compute it in SM. Using the building blocks \eqref{f-p-besq}, we distinguish whether the walker goes beyond $-m_1$ or not when hitting $k_2$ (see Fig.~\ref{fig:aged-rk}): 
\begin{align}
\label{full-p-A-laplace}
&\hat{P}_{\mathrm{incr}}(s)
=
\int_{-m_1}^{k_1}\! d\ell
\int_0^\infty\! da\;
\hat{P}^{k_2-k_1}_{2\beta}(a,s\!\mid\!0)\,
\hat{F}^{k_1-\ell}_{0}(s\!\mid\! a)
 \quad +\nonumber\\
&\iiint_0^\infty\!\!\!
\hat{P}^{k_2-k_1}_{2\beta}(a,s\!\mid\!0)\,
\hat{P}^{k_1+m_1}_{0}(b,s\!\mid\! a)\,
\hat{F}^{x}_{2-2\alpha}(s\!\mid\! b)\, da\,db\,dx.
\end{align}
The integration variables $\ell, a, b$ appearing in Eq.~\eqref{full-p-A-laplace} are shown on Fig~\ref{fig:aged-rk}. 
Remarkably, the Laplace-space expression \eqref{full-p-A-laplace} can be computed exactly (see Eq.~(19) of SM). 
The aged Ray–Knight representation thus provides a physically interpretable way to compute aged FPTs. Importantly, it relies only on the saturation property and therefore applies to the entire universality class of saturating SIRWs. For completeness, we also present in SM an alternative computation of \eqref{full-p-A-laplace} based on a direct matrix-algebraic formulation for pure SATW, independently validating our main result Eq.~\eqref{psi}.

Finally, we have obtained all the ingredients needed to compute aged FPTs for saturating SIRWs Eq.~\eqref{sktau-final}, which enable an exact expression for the quantity $\mathbb{P}(T_{k_2}-T_{k_1} \geq \tau)$ given in SM, where we also show how the appropriate record age limit $k_2=k_1+1=k+1$ leads to our more explicit main result Eq.~\eqref{psi}. Our approach relies on saturation, which renders the dynamics effectively Markovian in the bulk and confines memory to the boundaries; extending these results to non-saturating SIRWs, where memory is present in the entire visited interval, is an interesting open direction where the methodology presented in this Letter could be of use.

\emph{Acknowledgments.} We thank Léo Régnier for interesting discussions and for providing some numerical simulations of SATW.

Numerical simulation results as well as a Mathematica notebook for numerical evaluation of \eqref{psi} are available on \href{https://github.com/julien-bremont/records-satw/}{the following GitHub page}.

\clearpage



\section*{Supplementary material}

# Supplementary Material : Aging Record Statistics in Saturating Self-Interacting Random Walks

J. Brémont,[1] R. Voituriez,[2,3] and O. Bénichou[2]

[1]Collège de France, 3 Rue d'Ulm, 75005 Paris, France
[2]Laboratoire de Physique Théorique de la Matière Condensée, CNRS/Sorbonne Université, 4 Place Jussieu, 75005 Paris, France
[3]Laboratoire Jean Perrin, CNRS/Sorbonne Université, 4 Place Jussieu, 75005 Paris, France



## I. DISTRIBUTION OF THE AREA UNDER A SQUARED BESSEL PROCESS

In this section, we derive explicit expressions for the distribution of the area under a BESQ process that are required in our computations of relevant observables for SIRWs. To the best of our knowledge, some of these results do not appear in the existing literature, including in comprehensive references such as [1]: this motivates their presentation here.

We consider a $\mathrm{BESQ}_\delta$ process $(X_t)_{t\geq 0}$ starting from $X_0 = a$, and compute the joint Laplace-transformed distributions of the following quantities:

1. The joint distribution of the area under the process $\int_0^t X_u\,du$, and the position $X_t$ at time $t$:

$$\mathbb{E}_\delta^a\left[e^{-s\int_0^t X_u\,du}\,;\,X_t=b\right] = \int_{X_0=a,\,X_t=b}\mathcal{D}X\,e^{-s\int_0^t X_u\,du}. \tag{1}$$

2. The joint distribution of the area $\int_0^t X_u\,du$ and the first-passage time $T_0$ to the origin:

$$\mathbb{E}_\delta^a\left[e^{-s\int_0^t X_u\,du}\,;\,T_0=t\right] = \int_{X_0=a,\,T_0=t}\mathcal{D}X\,e^{-s\int_0^t X_u\,du}. \tag{2}$$

In both Eqs. (1) and (2), the path integral measure $\mathcal{D}X$ corresponds to that of a $\mathrm{BESQ}_\delta$ process: trajectories are weighted according to the law of a $\mathrm{BESQ}_\delta$ process over the interval $[0,t]$. In Eq. (1), the integration is carried out over all paths that reach $X_t = b$ at time $t$. In contrast, Eq. (2) involves integration over all positive trajectories with their first-passage to zero $T_0$ occurring exactly at time $t$. We stress that the normalization is chosen such that, when setting $s=0$ in (1), one recovers the propagator

$$\mathbb{E}_\delta^a\left[1\,;\,X_t=b\right] = \mathbb{P}(X_t=b\,|\,X_0=a), \tag{3}$$

while setting $s=0$ in (2) recovers the FPT density

$$\mathbb{E}_\delta^a\left[1\,;\,T_0=t\right] = \mathbb{P}(\text{FPT to 0 is at } t). \tag{4}$$

Fortunately, the first functional (1) can be computed using a result from [1] (formula 1.9.7, p. 384). Defining $\mu \equiv \sqrt{2s}$, the Laplace transform of the joint distribution of the area and the endpoint of a $\mathrm{BESQ}_\delta$ process reflected at zero is given by:

$$\mathbb{E}_\delta^a\left[e^{-s\int_0^t X_u\,du}\,;\,X_t=b\right] = \frac{\mu\left(\frac{b}{a}\right)^{\frac{\delta}{4}-\frac{1}{2}}e^{-\frac{1}{2}\mu(a+b)\coth(\mu t)}I_{\frac{\delta}{2}-1}\left(\frac{\mu\sqrt{ab}}{\sinh(\mu t)}\right)}{2\sinh(\mu t)} \equiv \hat{P}_\delta(b\mid a,t,s), \tag{5}$$

where $I_\nu$ denotes the modified Bessel function of the first kind. In the important case $a=0$, this becomes

$$\mathbb{E}_\delta^0\left[e^{-s\int_0^t X_u\,du}\,;\,X_t=b\right] = \frac{e^{\frac{-\mu b}{2}\coth(\mu t)}\left(\frac{\mu b}{2\sinh(\mu t)}\right)^{\delta/2}}{b\Gamma\left(\frac{\delta}{2}\right)} \tag{6}$$

The second functional (2), however, does not appear explicitly in the literature. We are only interested in the case $\delta<2$, as for $\delta\geq 2$, the $\mathrm{BESQ}_\delta$ process almost surely never hits the origin [2]. To evaluate this functional, we invoke

a classical time-reversal (or *duality*) argument [3]. Suppose the dimension satisfies $0 < \delta < 2$. In this regime, the $\mathrm{BESQ}_\delta$ process hits zero almost surely. In contrast, $4-\delta > 2$, so the $\mathrm{BESQ}_{4-\delta}$ process almost surely avoids zero. A key duality then holds: a $\mathrm{BESQ}_\delta$ process starting from $a > 0$ restricted (*not* conditioned, see (7)) to reaching $X_t = b$ without being absorbed before time $t$ (i.e., restricted to the event $\{T_0 > t\}$, where $T_0$ denotes the hitting time of zero) has the same law as a time-reversed $\mathrm{BESQ}_{4-\delta}$ process started from $b$, restricted to reaching $X_t = a$. More precisely, the following probabilities are equal:

$$\mathbb{P}^a_{\delta,\mathrm{abs}}(X_t = b)\, db = \mathbb{P}^a_\delta(X_t = b,\, T_0 > t)\, db = \mathbb{P}^b_{4-\delta}(X_t = a)\, da. \tag{7}$$

In fact, this duality (7) extends beyond the one-point marginal to arbitrary functionals of the process. In particular, we obtain the identity:

$$\mathbb{E}^a_{\delta,\mathrm{abs}}\left[e^{-s\int_0^t X_u\,du}\,;\, X_t = b\right] = \mathbb{E}^b_{4-\delta}\left[e^{-s\int_0^t X_u\,du}\,;\, X_t = a\right]. \tag{8}$$

To obtain the second functional (2), we now isolate the contribution of trajectories that hit zero for the first time in the infinitesimal time window $[t, t+dt]$. Using standard path integral reasoning, this gives:

$$\mathbb{E}^a_\delta\left[e^{-s\int_0^t X_u\,du}\,;\, T_0 = t\right] dt = \lim_{b\to 0} \mathbb{E}^a_{\delta,\mathrm{abs}}\left[e^{-s\int_0^t X_u\,du}\,;\, X_t = b\right] db. \tag{9}$$

To proceed, we need to determine how the $\mathrm{BESQ}_\delta$ process behaves as it approaches the origin. Recall that the $\mathrm{BESQ}_\delta$ process satisfies the stochastic differential equation:

$$dX_t = \delta\, dt + 2\sqrt{X_t}\, dB_t. \tag{10}$$

When $X_t = b$ is small, the term $2\sqrt{X_t}\, dB_t$ becomes negligible compared to the deterministic drift. Thus, to leading order, $db \approx \delta\, dt$. Inserting this into the previous expression, we obtain:

$$\mathbb{E}^a_\delta\left[e^{-s\int_0^t X_u\,du}\,;\, T_0 = t\right] = \delta \lim_{b\to 0} \mathbb{E}^a_{\delta,\mathrm{abs}}\left[e^{-s\int_0^t X_u\,du}\,;\, X_t = b\right]. \tag{11}$$

Finally, applying the time-reversal identity (8) and evaluating the limit using the expression (5), we obtain the joint Laplace transform of the area and the first-passage time for a BESQ process of dimension $2-\delta$:

$$\mathbb{E}^a_{2-\delta}\left[e^{-s\int_0^t X_u\,du}\,;\, T_0 = t\right] = \frac{\mu\, e^{-\frac{1}{2}\mu a \coth(\mu t)} \left(\frac{\mu a}{2\sinh(\mu t)}\right)^{\delta/2}}{\Gamma\left(\frac{\delta}{2}\right)\sinh(\mu t)} \equiv \hat{F}_{2-\delta}(0 \mid a, t, s), \tag{12}$$

where, as before, $\mu = \sqrt{2s}$. To the best of our knowledge, the only formula in the literature resembling (12) is equation (2.9.3) on p. 409 of [1]. However, that expression is given in the Laplace domain with respect to $t$, and no explicit inversion is provided. Moreover, the Laplace transform seems much too complicated to directly find the closed form (12) for its inverse. We verified numerically that our explicit result (12) agrees with the Laplace-inverted form of the expression given in [1].

## II. JOINT DISTRIBUTION OF MINIMUM $-m$ AND TIME $T_k$ TO THE MAXIMUM $k$

### A. Earlier results

Define $q_+(k, -m, t)$ (resp. $q_-(k, -m, t)$) as the joint distribution of the random variables $m$ and $t$ (resp. $k$ and $t$), where $t = T_k$ (resp. $t = T_{-m}$) is the first hitting time of site $k$ (resp. $-m$), and $-m$ (resp. $k$) is the minimum (resp. maximum) of the process $(X_s)_{s\le t}$. This distribution is of intrinsic interest and was computed in [4] to be:

$$\hat{q}_+(k, m, s) = \frac{1}{B(\alpha,\beta)} \cdot \frac{\mu \sinh^\alpha(k\mu)\sinh^{\beta-1}(\mu m)}{\sinh^{\alpha+\beta}(\mu(k+m))}, \quad \mu \equiv \sqrt{2s}. \tag{13}$$

Setting $s = 0$ in Eq. (13) yields the probability that the minimum is $-m$ when visiting $k$ for the first time:

$$\hat{q}_+(k, m, s = 0) = \frac{1}{B(\alpha,\beta)} \cdot \frac{k^\alpha m^{\beta-1}}{(k+m)^{\alpha+\beta}}. \tag{14}$$

### B. Aged exploration statistics

In this section, we derive the distribution of the time needed to extend a territory $[-m_1, \underline{k_1}]$ (where we underline the last site visited $k_1$) to territories $[-m_2, k_2]$. We distinguish the cases where the walker visits $k_2$ or $-m_2$ last.

#### 1. The walker last visits $k_2$

We now define the *aged* probability $q_+^{(2)}(k_2, m_2, t \,|\, k_1, m_1)$, which extend the one-time observable $q_+(k_1, m_1, t)$ to describe the statistics of a second excursion of the SIRW starting from a previously established visited territory.

Specifically, $q_+^{(2)}(k_2, m_2, t \,|\, k_1, m_1)$ denotes the joint probability density that the RW, having already reached site $k_1$ at time $T_{k_1}$ with span $[-m_1, k_1]$, subsequently reaches site $k_2 > k_1$ at time $T_{k_2} = T_{k_1} + t$, and that the new span at this time is $[-m_2, k_2]$:

$$q_+^{(2)}(k_2, m_2, t \,|\, k_1, m_1) = \mathbb{P}\left(T_{k_2} - T_{k_1} = t,\, \text{span} = [-m_2, k_2] \,|\, \text{span at } T_{k_1} = [-m_1, k_1]\right). \tag{15}$$

We stress that $q_+^{(2)}(k_2, m_2, t \,|\, k_1, m_1)$, as defined above, represents the joint distribution of the pair $(t, m_2)$: in turn, the variables $k_1, m_1, k_2$ are not random but conditionned on.

Using the aged Ray–Knight results, we will compute for $\hat{q}_+^{(2)}(k_2, m_2, s \,|\, k_1, m_1)$ in Laplace space $t \to s$ according to the aged Ray-Knight result, Fig. 2 of the main text. We must distinguish two cases. In the first case $m_1 = m_2$, the SIRW has not gone beyond its past minimum $-m_1$ at time $T_{k_2}$, and we have:

$$\begin{aligned}\hat{q}_+^{(2)}(k_2, m_1, s \mid k_1, m_1) = \int_0^{k_1+m_1} dl \int_0^\infty da\, \mathbb{E}^0 \left[ e^{-s\int_0^x Y_{2\beta}(u)\,du} \,\middle|\, Y_{2\beta}(k_2 - k_1) = a \right] \\ \times\, \mathbb{E}^a \left[ e^{-s\int_0^\infty \tilde{Y}_0(u)\,du} \,\middle|\, T_0^{(0)} = l \right].\end{aligned} \tag{16}$$

This evaluates to:

$$\hat{q}_+^{(2)}(k_2, m_1, s \mid k_1, m_1) = \left( \frac{\sinh\left(\mu(k_1 + m_1)\right)}{\sinh\left(\mu(k_2 + m_1)\right)} \right)^\beta. \tag{17}$$

In the second case, when $m_2 > m_1$, the SIRW does explore new territory to the left, and we write:

$$\begin{aligned}\hat{q}_+^{(2)}(k_2, m_2, s \,|\, k_1, m_1) = \iint_0^\infty & \mathbb{E}^0 \left[ e^{-s\int_0^x Y_{2\beta}(u)\,du} ;\, Y_{2\beta}(k_2 - k_1) = a \right] \\ & \times \mathbb{E}^a \left[ e^{-s\int_0^{k_1+m_1} \tilde{Y}_0(u)\,du} ;\, \tilde{Y}_0(k_1 + m_1) = b \right] \\ & \times \mathbb{E}^b \left[ e^{-s\int_0^\infty \widetilde{Y}_{2-2\alpha}(u)\,du} ;\, T_0^{(2-2\alpha)} = m_2 - m_1 \right] da\, db.\end{aligned} \tag{18}$$

This yields, again using Mathematica and unifying both cases:

$$\begin{aligned}\hat{q}_+^{(2)}(k_2, m_2, s \mid k_1, m_1) = \alpha\beta\mu \cdot & \frac{\sinh^{\beta-1}\big(\mu(k_1 + m_1)\big)}{\sinh^{\alpha+1}\big(\mu(k_2 + m_2)\big)} \cdot \sinh\big(\mu(k_2 - k_1)\big) \\ & \times \sinh^{\alpha-\beta}\big(\mu(k_2 + m_1)\big) \\ & \times {}_2F_1\left( \alpha + 1, 1 - \beta; 2; -\frac{\sinh\big(\mu(k_2 - k_1)\big) \cdot \sinh\big(\mu(m_2 - m_1)\big)}{\sinh\big(\mu(k_1 + m_1)\big) \cdot \sinh\big(\mu(k_2 + m_2)\big)} \right) \\ & + \delta(m_2 - m_1) \left( \frac{\sinh\left(\mu(k_1 + m_1)\right)}{\sinh\left(\mu(k_2 + m_1)\right)} \right)^\beta.\end{aligned} \tag{19}$$

#### 2. The walker last visits $-m_2$

Analogously (although this is not used in the main text) we define the aged probability $q_-^{(2)}(k_2, m_2, t \,|\, k_1, m_1)$ for the case where the walk continues after time $T_{k_1}$ and instead reaches a more negative site $-m_2 < -m_1$:

$$q_-^{(2)}(k_2, m_2, t \,|\, k_1, m_1) = \mathbb{P}\left(T_{-m_2} - T_{k_1} = t,\, \text{span} = [-m_2, k_2] \,|\, \text{span at } T_{k_1} = [-m_1, k_1]\right). \tag{20}$$

Thus, $q_-^{(2)}(k_2, m_2, t \,|\, k_1, m_1)$ as defined above defines the joint distribution of $(t, k_2)$.

These aged observables capture the evolution of the span conditioned on the past, and reflect the aging structure of the $\mathrm{SATW}_{\alpha,\beta}$. They serve as building blocks for two-time functionals and the analysis of memory and renewal effects in SIRWs.

Namely, using again aged Ray–Knight theory [5], we have

$$\begin{aligned}\hat{q}_-^{(2)}(k_2, m_2, s \,|\, k_1, m_1) = \iint_0^\infty & \mathbb{E}^0 \left[ e^{-s \int_0^x Y_{2\alpha}(u)\, du} \,;\, Y_{2\alpha}(m_2 - m_1) = a \right] \\ & \times \mathbb{E}^a \left[ e^{-s \int_0^{k_1+m_1} Y_2(u)\, du} \,;\, Y_2(k_1 + m_1) = b \right] \\ & \times \mathbb{E}^b \left[ e^{-s \int_0^\infty \widetilde{Y}_{2-2\beta}(u)\, du} \,;\, T_0^{(2-2\beta)} = k_2 - k_1 \right] da\, db. \end{aligned} \tag{21}$$

This integral can also be computed exactly using Mathematica. Setting $\mu = \sqrt{2s}$, we obtain

$$\begin{aligned}\hat{q}_-^{(2)}(k_2, m_2, s \mid k_1, m_1) = \beta\mu \cdot & \frac{\sinh^{\alpha-1}\big(\mu(k_1+m_1)\big)}{\sinh^{\beta+1}\big(\mu(k_2+m_2)\big)} \cdot \sinh^{-\alpha+\beta+1}\big(\mu(k_1+m_2)\big) \\ & \times {}_2F_1\left(1-\alpha, \beta+1; 1; -\frac{\sinh\big(\mu(k_2-k_1)\big)\cdot\sinh\big(\mu(m_2-m_1)\big)}{\sinh\big(\mu(k_1+m_1)\big)\cdot\sinh\big(\mu(k_2+m_2)\big)}\right). \end{aligned} \tag{22}$$

## III. COMPUTING $S_k(\tau)$

### A. Joint distribution of first-passage times to several targets

For generality, we first consider the joint distribution of two first-passage times (FPTs) to targets $k$ and $x$ with $k < x$. We introduce

$$\tilde{S}_x^k(t,\tau) \equiv \mathbb{P}(T_k = t,\ T_x = t+\tau)\,, \tag{23}$$

which describes the probability that the walker reaches $k$ for the first time at time $t$, and subsequently reaches $x$ for the first time at time $t+\tau$.

To compute $\tilde{S}_x^k(t,\tau)$, we condition on the minima reached before the respective first-passage events. Let $-m_1$ denote the minimum reached up to time $t$ (when $k$ is first visited), and $-m_2 \le -m_1$ the minimum reached up to time $t+\tau$ (when $x$ is first visited). Integrating over all admissible values of $m_1$ and $m_2$, we obtain

$$\tilde{S}_x^k(t,\tau) = \int_0^\infty dm_1 \int_{m_1}^\infty dm_2\, q_+(k, m_1, t)\, q_+^{(2)}(x, m_2, \tau \mid k, m_1), \tag{24}$$

where $q_+$, $q_+^{(2)}$ have been introduced above.

Taking the Laplace transform with respect to both $t$ and $\tau$ yields

$$\hat{\hat{\tilde{S}}}{}_x^k(s_1, s_2) = \int_0^\infty dm_1 \int_{m_1}^\infty dm_2\, \hat{q}_+(k, m_1, s_1)\, \hat{q}_+^{(2)}(x, m_2, s_2 \mid k, m_1). \tag{25}$$

In principle, Eq. (25), together with the explicit expressions of the observables $q_+$ and $q_+^{(2)}$ (see Eqs. (13) and (19)), fully determines the joint distribution of FPTs to arbitrary targets for the SATW. We recall that correlations between these FPTs are nontrivial, because the walker is non-Markovian and subsequent FPT statistics heavily depend on the visited territory when first reaching an earlier target. In contrast, they are simply handled in the case of renewal processes. Finally, the method straightforwardly generalizes to joint correlations of FPTs at $N$ different targets $0 < x_1 < \cdots < x_N$.

In the present work, we focus on the case $s_1 = 0$ and on the near-target regime $|x-k| \ll k$, which is the appropriate limit to study the record ages $\tau_k$ of interest in the Letter. The record-age distribution $S_k(\tau)$ considered in the main text is obtained as the particular case $x = k+1$, namely

$$S_k(\tau) = \mathbb{P}(\tau_k \ge \tau) = \int_0^\infty dt \int_\tau^\infty d\tau'\, \tilde{S}_{k+1}^k(t, \tau'). \tag{26}$$

Using the identity

$$\int_0^\infty e^{-s\tau} d\tau \int_\tau^\infty f(\tau')\, d\tau' = \frac{1}{s}\left(\hat{f}(0) - \hat{f}(s)\right), \tag{27}$$

we obtain the explicit representation

$$\int_0^\infty e^{-s\tau} S_k(\tau)\, d\tau = \frac{1}{s}\left(1 - \int_0^\infty dm_1 \int_{m_1}^\infty dm_2\, \hat{q}_+(k, m_1, 0)\, \hat{q}_+^{(2)}(k+1, m_2, s \mid k, m_1)\right). \tag{28}$$

We now compute this expression explicitly in the following, showing how to invert the Laplace transform to obtain an exact expression for $S_k(\tau)$.

### B. Explicit distribution of the record ages

In this section we compute $S_k(\tau) \equiv \mathbb{P}(\tau_k \geq \tau)$ for generic $(\alpha, \beta)$: we start from the Laplace representation Eq. (25), insert the explicit formulas for $\hat{q}_+$ and $\hat{q}_+^{(2)}$, expand regime $k+1 = k + dk$ with $\mu dk \ll 1$ (where $\mu = \sqrt{2s}$), and finally perform the $m_2$-integration.

*a. Starting point.* From Eq. (25) we have

$$\int_0^\infty e^{-s\tau} S_k(\tau)\, d\tau = \frac{1}{s}\left(1 - \int_0^\infty dm_1 \int_{m_1}^\infty dm_2\, \hat{q}_+(k, m_1, 0)\, \hat{q}_+^{(2)}(k+1, m_2, s \,|\, k, m_1)\right). \tag{29}$$

We now set $(k_1, k_2) = (k, k + dk)$ and work at first order in $\mu dk \ll 1$.

*b. Step 1: expansion for $\mu dk \ll 1$.* The Dirac part from Eq. (19) gives

$$\left(\frac{\sinh(\mu(k+m))}{\sinh(\mu(k+dk+m))}\right)^\beta = 1 - \beta\, \mu\, dk\, \coth\big(\mu(k+m)\big) + O((\mu dk)^2). \tag{30}$$

For the continuous part we use

$$\sinh(\mu dk) = \mu dk + O((\mu dk)^3), \qquad \sinh^{\alpha-\beta}(\mu(k+dk+m_1)) = \sinh^{\alpha-\beta}(\mu(k+m_1)) + O(\mu dk),$$

and we note that the hypergeometric argument in (19) is $O(\mu dk)$, hence

$${}_2F_1(\alpha+1, 1-\beta; 2; -O(\mu dk)) = 1 + O(\mu dk),$$

which contributes only at order $O((\mu dk)^2)$ once multiplied by $\sinh(\mu dk) = O(\mu dk)$, and can thus be dropped at the present order. With these simplifications, the continuous part of (19) reduces at leading order to

$$\hat{q}^{(2)}_{+,\,\mathrm{cont}}(k+dk, m_2, s \,|\, k, m_1) = \alpha\beta\, \mu^2\, dk\, \frac{\sinh^{\beta-1}\big(\mu(k+m_1)\big)\, \sinh^{\alpha-\beta}\big(\mu(k+m_1)\big)}{\sinh^{\alpha+1}\big(\mu(k+m_2)\big)} + O((\mu dk)^2). \tag{31}$$

Since $(\beta - 1) + (\alpha - \beta) = \alpha - 1$, we obtain the key simplification

$$\hat{q}^{(2)}_{+,\,\mathrm{cont}}(k+dk, m_2, s \,|\, k, m_1) = \alpha\beta\, \mu^2\, dk\, \frac{\sinh^{\alpha-1}\big(\mu(k+m_1)\big)}{\sinh^{\alpha+1}\big(\mu(k+m_2)\big)} + O((\mu dk)^2). \tag{32}$$

Inserting (30) and (32) into (29) yields, at leading order in $\mu dk$,

$$\begin{aligned} 1 - s\hat{S}_k(s) \underset{\mu dk \ll 1}{\sim} 1 &- dk\, \frac{\beta\mu}{B(\alpha,\beta)} \int_0^\infty dm\, \frac{k^\alpha m^{\beta-1}}{(k+m)^{\alpha+\beta}} \coth\big(\mu(k+m)\big) \\ &+ dk\, \frac{\alpha\beta\mu^2}{B(\alpha,\beta)} \int_0^\infty dm_1 \int_{m_1}^\infty dm_2\, \frac{k^\alpha m_1^{\beta-1}}{(k+m_1)^{\alpha+\beta}}\, \frac{\sinh^{\alpha-1}\big(\mu(k+m_1)\big)}{\sinh^{\alpha+1}\big(\mu(k+m_2)\big)}. \end{aligned} \tag{33}$$

*c. Step 2: integrating over $m_2$.* We use the integral identity:

$$\int_{m_1}^{\infty} \frac{dm_2}{\sinh^{\alpha+1}\big(\mu(k+m_2)\big)} = \frac{1}{\mu}\int_0^{\infty} \frac{da}{\sinh^{\alpha+1}\big(\mu(k+m_1)+a\big)}$$
$$= \frac{1}{\mu}\,\frac{e^{-\mu(k+m_1)}}{\sinh^{\alpha}\big(\mu(k+m_1)\big)}\,\frac{2}{\alpha+1}\,{}_2F_1\!\left(\frac{1-\alpha}{2},1;\frac{3+\alpha}{2};e^{-2\mu(k+m_1)}\right). \tag{34}$$

Multiplying (34) by $\sinh^{\alpha-1}(\mu(k+m_1))$ gives a single sinh in the denominator:

$$\sinh^{\alpha-1}(\mu(k+m_1)) \times \frac{1}{\sinh^{\alpha}(\mu(k+m_1))} = \frac{1}{\sinh(\mu(k+m_1))}.$$

Therefore the second (double) integral in (33) becomes

$$dk\,\frac{2\alpha\beta\mu}{B(\alpha,\beta)(\alpha+1)}\int_0^{\infty} dm\,\frac{k^{\alpha}m^{\beta-1}}{(k+m)^{\alpha+\beta}}\,\frac{e^{-\mu(k+m)}}{\sinh\big(\mu(k+m)\big)}\,{}_2F_1\!\left(\frac{1-\alpha}{2},1;\frac{3+\alpha}{2};e^{-2\mu(k+m)}\right). \tag{35}$$

*d. Step 3: isolating the k-independent part.* Using $\coth(x) = 1 + \frac{e^{-x}}{\sinh(x)}$ and the normalization

$$\int_0^{\infty} dm\,\frac{k^{\alpha}m^{\beta-1}}{(k+m)^{\alpha+\beta}} = B(\alpha,\beta), \tag{36}$$

we rewrite the first integral in (33) as a sum of a constant term and an $e^{-\mu(k+m)}/\sinh(\mu(k+m))$ term. Combining it with (35) yields the compact expression

$$s\hat{S}_k(s) = \beta\,\mu\,dk\Bigg(1 - \frac{1}{B(\alpha,\beta)}\int_0^{\infty} dm\,\frac{k^{\alpha}m^{\beta-1}}{(k+m)^{\alpha+\beta}}\,\frac{e^{-\mu(k+m)}}{\sinh\big(\mu(k+m)\big)}$$
$$\times\left[\frac{2\alpha}{\alpha+1}\,{}_2F_1\!\left(\frac{1-\alpha}{2},1;\frac{3+\alpha}{2};e^{-2\mu(k+m)}\right) - 1\right]\Bigg). \tag{37}$$

*e. Step 4.* We now simplify the bracket in (37) using a contiguous relation for ${}_2F_1$. One finds the identity (valid for all $\alpha > 0$ by analytic continuation)

$$\frac{e^{-u}}{\sinh(u)}\left[\frac{2\alpha}{\alpha+1}\,{}_2F_1\!\left(\frac{1-\alpha}{2},1;\frac{3+\alpha}{2};e^{-2u}\right) - 1\right] = 2\,\frac{\alpha-1}{\alpha+1}\,e^{-2u}\,{}_2F_1\!\left(\frac{3-\alpha}{2},1;\frac{3+\alpha}{2};e^{-2u}\right), \qquad u > 0. \tag{38}$$

This allows us to write the kernel $K_{\alpha,\beta}(\mu(k+m))$ which multiplies the term $\frac{k^{\alpha}m^{\beta-1}}{(k+m)^{\alpha+\beta}}$ in (37):

$$\boxed{K_{\alpha,\beta}(u) \equiv 2\,\frac{\alpha-1}{\alpha+1}\,e^{-2u}\,{}_2F_1\!\left(\frac{3-\alpha}{2},1;\frac{3+\alpha}{2};e^{-2u}\right)} \qquad \text{(note: } K_{\alpha,\beta} \text{ depends on } \alpha \text{ only).} \tag{39}$$

With this notation, (37) becomes

$$s\hat{S}_k(s) = \beta\,\mu\,dk\left(1 - \frac{1}{B(\alpha,\beta)}\int_0^{\infty} dm\,\frac{k^{\alpha}m^{\beta-1}}{(k+m)^{\alpha+\beta}}\,K_{\alpha,\beta}\big(\mu(k+m)\big)\right). \tag{40}$$

*f. Step 5: scaling form for Laplace inversion.* Changing variables $m = ky$ yields the expression

$$\hat{S}_k(s) = \sqrt{\frac{2}{s}}\,\beta\,dk\left(1 - \frac{1}{B(\alpha,\beta)}\int_0^{\infty} dy\,\frac{y^{\beta-1}}{(1+y)^{\alpha+\beta}}\,K_{\alpha,\beta}\big(\mu k(1+y)\big)\right), \qquad \mu = \sqrt{2s}. \tag{41}$$

*g. Step 6: exact τ-space inversion (asymmetric case).* We now invert (41) in closed form.
Using the explicit kernel

$$K_{\alpha,\beta}(u) = 2\,\frac{\alpha-1}{\alpha+1}\,e^{-2u}\,{}_2F_1\!\left(\frac{3-\alpha}{2},1;\frac{3+\alpha}{2};e^{-2u}\right), \qquad u > 0,$$

we expand the hypergeometric function into its power series

$$ {}_2F_1(a,1;c;z) = \sum_{n=0}^{\infty} \frac{(a)_n}{(c)_n}\, z^n, \qquad |z|<1, $$

which yields the convergent exponential series representation

$$ K_{\alpha,\beta}(u) = 2\,\frac{\alpha-1}{\alpha+1} \sum_{n=0}^{\infty} \frac{\left(\frac{3-\alpha}{2}\right)_n}{\left(\frac{3+\alpha}{2}\right)_n}\, e^{-2(n+1)u}. \tag{42} $$

Inserting (42) into (41) and exchanging sum and integral, we obtain

$$ \hat{S}_k(s) = \sqrt{\frac{2}{s}}\,\beta\, dk \left(1 - \frac{2(\alpha-1)}{(\alpha+1)B(\alpha,\beta)} \sum_{n=0}^{\infty} \frac{\left(\frac{3-\alpha}{2}\right)_n}{\left(\frac{3+\alpha}{2}\right)_n} \int_0^{\infty} dy\, \frac{y^{\beta-1}}{(1+y)^{\alpha+\beta}}\, e^{-2(n+1)\mu k(1+y)}\right), \qquad \mu = \sqrt{2s}. \tag{43} $$

The Laplace inversion is then performed term-by-term using the standard identity

$$ \mathcal{L}^{-1}_{s\to\tau}\left[\frac{1}{\sqrt{s}}\, e^{-c\sqrt{s}}\right] = \frac{1}{\sqrt{\pi\tau}}\, \exp\!\left(-\frac{c^2}{4\tau}\right), \qquad c>0, \tag{44} $$

along with the fact that

$$ e^{-2(n+1)\mu k(1+y)} = \exp\!\left(-2(n+1)k(1+y)\sqrt{2s}\right) = \exp\!\left(-c\sqrt{s}\right), \qquad c = 2\sqrt{2}\,(n+1)k(1+y). $$

Applying (44) to each term of (43) yields the exact expression in physical time $\tau$, valid in the scaling regime $k^{-2}\tau$ fixed with $\tau, k\to\infty$:

$$ \boxed{S_k(\tau) = \beta\sqrt{\frac{2}{\pi\tau}}\left(1 - \frac{2(\alpha-1)}{(\alpha+1)B(\alpha,\beta)} \sum_{n=0}^{\infty} \frac{\left(\frac{3-\alpha}{2}\right)_n}{\left(\frac{3+\alpha}{2}\right)_n} \int_0^{\infty} dy\, \frac{y^{\beta-1}}{(1+y)^{\alpha+\beta}}\, \exp\!\left[-\frac{2(n+1)^2k^2(1+y)^2}{\tau}\right]\right),} \tag{45} $$

where we set $dk = 1$ for the increment in record index.

### C. Asymptotics of the scaling function

From the exact result (45), we directly identify the scaling function $\psi$ as

$$ \psi\left(x = \frac{\tau}{k^2}\right) = \beta\sqrt{\frac{2}{\pi x}}\left(1 - \frac{2(\alpha-1)}{(\alpha+1)B(\alpha,\beta)} \sum_{n=0}^{\infty} \frac{\left(\frac{3-\alpha}{2}\right)_n}{\left(\frac{3+\alpha}{2}\right)_n} \int_0^{\infty} dy\, \frac{y^{\beta-1}}{(1+y)^{\alpha+\beta}}\, \exp\!\left[-\frac{2(n+1)^2(1+y)^2}{x}\right]\right). \tag{46} $$

From (46), it is clear that in the small-$x$ regime, the scaling function simplifies to:

$$ \psi(x) \underset{x\ll 1}{\sim} \beta\sqrt{\frac{2}{\pi x}}. \tag{47} $$

The large-time regime $x \gg 1$ is more intricate. We focus here on the range $1<\alpha<2$ to make computations explicit, and later extend the result to all $\alpha>0$ via analytic continuation, supported numerically Fig. 2 of the main text. We first remark that the following hypergeometric series is actually unity for $1<\alpha<2$:

$$ \frac{2(\alpha-1)}{\alpha+1} \sum_{n=0}^{\infty} \frac{\left(\frac{3-\alpha}{2}\right)_n}{\left(\frac{3+\alpha}{2}\right)_n} = 1. \tag{48} $$

Using (48) and the integral representation of the Beta function $B(\alpha,\beta) = \int_0^\infty dy\, \frac{y^{\beta-1}}{(1+y)^{\alpha+\beta}}$, we obtain the following representation of $\psi(x)$ in the large-$x$ limit:

$$ \psi(x) \underset{x\gg 1}{\sim} \beta\sqrt{\frac{2}{\pi x}} \cdot \frac{2(\alpha-1)}{(\alpha+1)B(\alpha,\beta)} \sum_{n=0}^{\infty} \frac{\left(\frac{3-\alpha}{2}\right)_n}{\left(\frac{3+\alpha}{2}\right)_n} \int_0^{\infty} dy\, \frac{y^{\beta-1}}{(1+y)^{\alpha+\beta}} \left(1 - e^{-\frac{2(n+1)^2(1+y)^2}{x}}\right). \tag{49} $$

To extract the leading asymptotics, we convert the sum over $n$ into an integral using the substitution $n = \sqrt{x}u$ for large $x$. Using Stirling's approximation for the ratio of Pochhammer symbols, we find:

$$\frac{2(\alpha-1)}{\alpha+1} \cdot \frac{\left(\frac{3-\alpha}{2}\right)_n}{\left(\frac{3+\alpha}{2}\right)_n} \underset{n=\sqrt{x}u,\, x\to\infty}{\sim} -x^{-\alpha/2} \cdot \left( \frac{2u^{-\alpha}\Gamma\left(\frac{1+\alpha}{2}\right)}{\Gamma\left(\frac{1}{2}-\frac{\alpha}{2}\right)} \right). \tag{50}$$

Substituting into (49), we obtain:

$$\psi(x) \underset{x\gg 1}{\sim} -\beta\sqrt{\frac{2}{\pi x^{\alpha}}} \cdot \frac{2}{B(\alpha,\beta)} \cdot \frac{\Gamma\left(\frac{1+\alpha}{2}\right)}{\Gamma\left(\frac{1}{2}-\frac{\alpha}{2}\right)} \int_0^\infty u^{-\alpha} du \int_0^\infty dy \, \frac{y^{\beta-1}}{(1+y)^{\alpha+\beta}} \left(1 - e^{-2u^2(1+y)^2}\right). \tag{51}$$

Both integrals can be evaluated explicitly for $1 < \alpha < 2$ using standard identities [6]:

$$\int_0^\infty \left(1 - e^{-2u^2(1+y)^2}\right) u^{-\alpha} du = -2^{\frac{\alpha-3}{2}} (1+y)^{\alpha-1} \Gamma\left(\frac{1}{2} - \frac{\alpha}{2}\right),$$

$$\int_0^\infty \frac{y^{\beta-1}(1+y)^{\alpha-1}}{(1+y)^{\alpha+\beta}} \, dy = \frac{1}{\beta}.$$

Putting everything together, we obtain the explicit large-$x$ asymptotics:

$$\psi(x) \underset{x\gg 1}{\sim} \left(\frac{2}{x}\right)^{\alpha/2} \cdot \frac{\Gamma\left(\frac{1+\alpha}{2}\right)}{\sqrt{\pi}\, B(\alpha,\beta)}. \tag{52}$$

This result confirms the predicted scaling behavior $\psi(x) \propto x^{-\alpha/2}$ and yields an explicit expression for the prefactor.

## IV. AN INDEPENDENT DERIVATION OF THE DISTRIBUTION OF THE AREA $\mathcal{A}$ UNDER THE LOCAL TIME INCREMENTS $I_x(k_2 \,|\, k_1, m_1)$

**Outline of the derivation.** Here, we present an independent derivation of the quantities $\hat{q}_{\pm}^{(2)}(s)$ given by Eqs. (19) and (22), which are needed for the computation of Eq. (20) of the main text, without invoking the aged Ray–Knight construction. This more direct approach, similar in spirit to the computation found in [4], is valuable as an independent validation of our main result and of its building blocks Eqs. (19) and (22).

The strategy is the following.

1. We first start with setting the Laplace variable $s = 0$, i.e. we consider only time-integrated quantities $\hat{q}_{\pm}^{(2)}(s = 0)$. We will consider the effect of nonzero $s$ at the end.

2. We derive functional relations for such probabilities.

3. These relations remarkably close for the ratio observable $\gamma = \frac{\hat{q}_{+}^{(2)}(k_2,m_2,s=0|k_1,m_1)}{\hat{q}_{-}^{(2)}(k_2,m_2,s=0|k_1,m_1)}$.

4. A joint continuum limit yields coupled partial differential equations.

5. These reduce to a Riccati equation for $\gamma$, admitting an exact hypergeometric solution.

6. We show how to reconstruct $\hat{q}_{\pm}^{(2)}(s = 0)$ from the knowledge of $\gamma$, recovering our results Eqs. (19) and (22).

For simplicity, we consider symmetric SATW, that is $\alpha = \beta \equiv \phi$.

### A. Functional equation and ratio observable

Thanks to translation invariance, we can assume that $m_1 = 0$, and $k_1 = L_1$ is the length of the already visited interval.

A first key observation is the following set of functional relations, akin to the similar one found in [4]

$$\begin{aligned}\hat{q}_+^{(2)}(k_2{+}1, m_2 \,|\, L_1) &= \hat{q}_+^{(2)}(k_2, m_2 \,|\, L_1)\, \mathbb{P}\big(k_2{+}1 \text{ reached before } -m_2{-}1 \,\big|\, [-m_2, \underline{k_2}]\big) \\ &\quad + \hat{q}_-^{(2)}(k_2, m_2 \,|\, L_1)\, \mathbb{P}\big(k_2{+}1 \text{ reached before } -m_2{-}1 \,\big|\, [\underline{-m_2}, k_2]\big) \qquad (53)\\ \hat{q}_-^{(2)}(k_2{+}1, m_2 \,|\, L_1) &= \hat{q}_+^{(2)}(k_2, m_2 \,|\, L_1)\, \mathbb{P}\big(-m_2{-}1 \text{ reached before } k_2{+}1 \,\big|\, [-m_2, \underline{k_2}]\big) \\ &\quad + \hat{q}_-^{(2)}(k_2, m_2 \,|\, L_1)\, \mathbb{P}\big(-m_2{-}1 \text{ reached before } k_2{+}1 \,\big|\, [\underline{-m_2}, k_2]\big)\,. \qquad (54)\end{aligned}$$

The splitting probabilities (53) can be computed easily, because the SATW behaves as a simple RW in its span, with transition probabilities only modified at the boundaries. Let us write

$$P_{+,+}(L) = \mathbb{P}\big(k_2{+}1 \text{ reached before } -m_2{-}1 \,\big|\, [-m_2, \underline{k_2}]\big) = P_{-,-}(L) \qquad (55)$$

$$P_{+,-}(L) = \mathbb{P}\big(k_2{+}1 \text{ reached before } -m_2{-}1 \,\big|\, [\underline{-m_2}, k_2]\big) = P_{-,+}(L). \qquad (56)$$

Then, we introduce the matrix of splitting probabilities

$$F_L(0) = \begin{pmatrix} P_{+,+}(L) & P_{+,-}(L) \\ P_{-,+}(L) & P_{-,-}(L) \end{pmatrix}$$

Introducing the ratio

$$\gamma(x,y) = \gamma = \frac{\hat{q}_+^{(2)}(k_2, m_2, s=0 \mid k_1, m_1)}{\hat{q}_-^{(2)}(k_2, m_2, s=0 \mid k_1, m_1)}, \qquad (57)$$

the solution to (53) reads

$$\begin{aligned}\hat{q}_+^{(2)}(L_1+x, y \,|\, L_1) &= F_{L_1}(-|+)F_{L_1+1}(-|-)\dots F_{L_1+y-1}(-|-)F_{L_1+y}(+|-)\prod_{j=1}^{x-1}\left[\frac{1}{\gamma(j,y)}\hat{F}_{L_1+y+j}(+|-) + \hat{F}_{L_1+y+j}(+|+)\right] \\ \hat{q}_-^{(2)}(L_1+x, y \,|\, L_1) &= F_{L_1}(+|+)F_{L_1+1}(+|+)\dots F_{L_1+x-1}(+|+)F_{L_1+x}(-|+)\prod_{j=1}^{y-1}\left[\gamma(x,j)\hat{F}_{L_1+x+j}(-|+) + \hat{F}_{L_1+x+j}(-|-)\right],\end{aligned} \qquad (58)$$

where we used that there is a unique path that leads to incrementing by 1 either the right (i.e. $L_1$) or the left starting boundary (i.e. 0):

$$\begin{aligned}\hat{q}_+^{(2)}(L_1+1, y \,|\, L_1) &= F_{L_1}(-|+)F_{L_1+1}(-|-)\dots F_{L_1+y-1}(-|-)F_{L_1+y}(+|-) \\ \hat{q}_-^{(2)}(L_1+x, 1 \,|\, L_1) &= F_{L_1}(+|+)F_{L_1+1}(+|+)\dots F_{L_1+x-1}(+|+)F_{L_1+x}(-|+).\end{aligned} \qquad (59)$$

In [4], we showed that the transfer matrices have the explicit expression

$$F_L(0) = \frac{1}{L+2\phi}\begin{pmatrix} L+\phi & \phi \\ \phi & L+\phi \end{pmatrix}. \qquad (60)$$

One therefore exact product expressions for the probabilities $q_\pm^{(2)}$:

$$\begin{aligned}\hat{q}_+^{(2)}(L_1+x, y \,|\, L_1) &= \phi^2\frac{\prod_{l=1}^{y-1}(L_1+l+\phi)\prod_{j=1}^{x-1}\big(\phi\gamma(j,y)^{-1} + L_1+j+y+\phi\big)}{\prod_{j=0}^{x+y-1}(L_1+j+2\phi)}, \\ \hat{q}_-^{(2)}(L_1+x, y \,|\, L_1) &= \phi\frac{\prod_{l=0}^{x-1}(L_1+l+\phi)\prod_{j=1}^{y-1}(\phi\gamma(x,j) + L_1+j+x+\phi)}{\prod_{j=0}^{x+y-1}(L_1+j+2\phi)}.\end{aligned} \qquad (61)$$

These expressions (61) are not yet explicit at this stage because one lacks the form of $\gamma$. This is the focus of the next sections.

While not obvious at this stage, we will see that these expressions (61) imply closed recurrence relations for $\gamma$. We make the self-consistent hypothesis that, for $y \geq x$, one has

$$\boxed{\gamma(x, y+1) = \frac{x+y+L_1+\phi(1+\gamma(y,x)^{-1})}{x+y+L_1+\phi(1+\gamma(x,y))}\frac{y}{y+1}\gamma(x,y)} \tag{62}$$

and

$$\boxed{\gamma(x+1, y) = \gamma(x,y)\frac{L_1+x+\phi+1}{L_1+x+\phi}\frac{x+y+L_1+\phi(1+\gamma(x,y)^{-1})}{x+y+L_1+\phi(1+\gamma(y-1,x+1))}.} \tag{63}$$

Both identities can be checked to hold for finite values of $x, y$ plugging in the exact expression of $F_L$ (60) using the matrix formulation (58). The following sections validate them in the scaling limit defined below.

### B. Continuum scaling limit

Exact solution of the discrete relations (61) appears intractable. We therefore consider a joint large-span limit in which both boundaries grow proportionally to the initial span:

$$x = z\Delta, \qquad y = (1-z)\Delta, \qquad \Delta = \delta L_1.$$

We introduce the scaling function

$$\gamma(x,y) = g\left(z = \frac{x}{x+y},\, \delta = \frac{x+y}{L_1}\right).$$

Because each update changes the total span, $g$ depends on both variables $z$ and $\delta$.
Taking $L_1 \to \infty$ in (61) crucially yields two coupled PDEs,

$$\delta\partial_\delta g - z\partial_z g = \Big(-\frac{1}{1-z} + \phi\,\frac{1-g(z,\delta)g(1-z,\delta)}{(1+\delta^{-1})g(1-z,\delta)}\Big)g, \tag{64}$$

$$\delta\partial_\delta g + (1-z)\partial_z g = \Big(\frac{1}{z+\delta^{-1}} + \phi\,\frac{1-g(z,\delta)g(1-z,\delta)}{(1+\delta^{-1})g(z,\delta)}\Big)g. \tag{65}$$

### C. Reduction to a Riccati equation

Remarkably, combining the two PDEs eliminates the reinforcement parameter $\phi$ and yields

$$\partial_z \log\frac{g(z,\delta)}{g(1-z,\delta)} = \frac{1}{z} + \frac{1}{1-z} + \frac{1}{z+\delta^{-1}} + \frac{1}{1-z+\delta^{-1}}. \tag{66}$$

Integration gives

$$\frac{g(z,\delta)}{g(1-z,\delta)} = \frac{z(\delta z+1)}{(1-z)(\delta(1-z)+1)}. \tag{67}$$

Substitution into the PDE (64) reduces the problem to a single Riccati equation,

$$\boxed{\partial_z g = \frac{1+\delta}{(1-z)(1+z\delta)}g + \delta\phi\frac{1-2z}{1-z}\left[\frac{1}{1+\delta(1-z)} - \frac{1-z}{z(1+z\delta)}g^2\right].} \tag{68}$$

### D. Exact solution

The Riccati equation (68) admits an exact solution,

$$\boxed{g(z,\delta) = \frac{z(z+\delta^{-1})}{(1-2z)\phi}\partial_z \log\left[{}_2F_1\left(-\phi,\phi;1;\frac{(z-1)z\delta^2}{\delta+1}\right)\right].} \tag{69}$$

Remarkably, the same hypergeometric structure as in (13) therefore emerges without invoking the aged Ray–Knight construction.

*1. Continuum limit of $\hat{\pi}_{x,\underline{-y}}$*

We start from the exact expression (61)

$$\hat{q}_{-}^{(2)}(L_1+x,y\,|\,L_1) = \phi\frac{\displaystyle\prod_{l=0}^{x-1}(L_1+l+\phi)\prod_{j=1}^{y-1}(\phi\gamma(x,j)+L_1+j+x+\phi)}{\displaystyle\prod_{j=0}^{x+y-1}(L_1+j+2\phi)}. \tag{70}$$

In contrast with the complementary contribution $\hat{q}_{+}^{(2)}(L_1+x,y\,|\,L_1)$, no Dirac peak at $x=0$ appears in this case.

We now consider the scaling regime

$$x+y=\Delta=\delta L_1, \qquad j=u\Delta, \qquad u\in[0,1-z].$$

*a. Continuum approximation of the product term.* Taking the logarithm of the second product yields

$$\log\prod_{j=1}^{y-1}\left[\gamma(x,j)F_{L_1+x+j}(+|-)+F_{L_1+x+j}(+|+)\right]. \tag{71}$$

Replacing sums by integrals using $j=u\Delta$ gives

$$\sim \delta L_1\int_{\Delta^{-1}}^{1-z}du\,\log\left[g\left(\frac{z}{z+u},\delta(z+u)\right)\frac{\phi}{L_1(1+\delta(z+u))}+1-\frac{\phi}{L_1(1+\delta(z+u))}\right]. \tag{72}$$

Expanding the logarithm for large $L_1$,

$$\sim \delta L_1\int_{\Delta^{-1}}^{1-z}du\,\log\left[1+\left(g\left(\frac{z}{z+u},\delta(z+u)\right)-1\right)\frac{\phi}{L_1(\delta(z+u)+1)}\right], \tag{73}$$

The lower bound may now safely be set to 0, since $\delta^{-1}$ regularizes the integral. At leading order, we obtain

$$\sim \delta\phi\int_0^{1-z}\left(g\left(\frac{z}{z+u},\delta(z+u)\right)-1\right)\frac{du}{\delta(z+u)+1}. \tag{74}$$

*b. Prefactor contribution.* The remaining product evaluates to

$$\frac{1}{L_1+x+2\phi}\prod_{l=0}^{x-1}\frac{L_1+l+\phi}{L_1+l+2\phi}. \tag{75}$$

Using Gamma functions,

$$=\frac{1}{L_1(1+\delta z)+2\phi}\frac{\Gamma(L_1+2\phi)\,\Gamma(z\delta L_1+L_1+\phi)}{\Gamma(L_1+\phi)\,\Gamma(z\delta L_1+L_1+2\phi)}, \tag{76}$$

and taking $L_1\to\infty$ gives

$$\sim\frac{1}{L_1}\,(1+\delta z)^{-\phi-1}. \tag{77}$$

Combining all contributions yields

$$\boxed{\hat{q}_{-}^{(2)}(L_1+x,y\,|\,L_1)\sim C(\delta,\phi)\frac{\delta}{(1+\delta z)^{\phi+1}}\exp\left[\delta\phi\int_0^{1-z}\left(g\left(\frac{z}{z+u},\delta(z+u)\right)-1\right)\frac{du}{\delta(z+u)+1}\right].} \tag{78}$$

*c. Exact evaluation.* Evaluating the integral using the explicit solution for $g$ provides the closed form

$$\boxed{\hat{q}_-^{(2)}(L_1+x,y\,|\,L_1)\sim\frac{\delta\phi}{(1+\delta)^\phi}\frac{\delta(1-z)+1}{1+\delta+\delta^2z(1-z)}\,{}_2F_1\left(-\phi,\phi;1;\frac{\delta^2(z-1)z}{\delta+1}\right).}\tag{79}$$

From (79) and (69), we directly deduce the expression of $\hat{q}_+^{(2)}$ as:

$$\boxed{\hat{q}_+^{(2)}(L_1+x,y\,|\,L_1)\sim\frac{\delta^2\phi^2z}{(\delta+1)^{\phi+1}}\,{}_2F_1\left(1-\phi,1+\phi;2;\frac{(z-1)z\delta^2}{\delta+1}\right)+\frac{\delta_{\text{Dirac}}(z-1)}{(1+\delta)^\phi}.}\tag{80}$$

We checked self-consistently that the integration constant $C(\delta,\phi)=\phi$ by normalization of the probability

$$1=\sum_{k=0}^{L}\left(\hat{q}_-^{(2)}(L_1+k,L-k\,|\,L_1)+\hat{q}_+^{(2)}(L_1+k,L-k\,|\,L_1)\right),\tag{81}$$

which simply expresses that the walker always visits $L+L_1$ sites starting from a visited territory of size $L_1$.

*d. Consistency checks.* For $\phi=1$ one recovers

$$\hat{q}_-^{(2)}(L_1+x,y\,|\,L_1)\sim\delta\,\frac{1+(1-z)\delta}{(1+\delta)^2},\tag{82}$$

using $g(z,\delta)=z/(1-z+\delta^{-1})$, in agreement with the Brownian limit obtained by a simple product of splitting probabilities.

Also, the expression (79) reproduces the non-aged result [7] in the limit $\delta\to\infty$.

### E. Time-dependent extension ($\mu\neq0$)

We now extend the previous analysis to the time–dependent problem. Assume that at some initial time $t_0$ the explored domain is $[-m,\underline{k}]$, and consider the probability that at a later time $t$ the walk reaches the enlarged span $[-m-y,\underline{k+x}]$. The framework is therefore identical to the previous section, with the additional Laplace parameter $\mu=\sqrt{2s}>0$.

The first–passage matrices are those obtained for SATW at finite $\mu$ (see Ref. [4]):

$$\hat{F}_L\sim\mathbb{1}_2+\mu\begin{bmatrix}-\phi\coth(L\mu) & \frac{\phi}{\sinh L\mu}\\ \frac{\phi}{\sinh L\mu} & -\phi\coth(L\mu)\end{bmatrix}\tag{83}$$

*a. Scaling limit.* We consider the joint limit that is relevant to us:

$$L_1\gg1,\qquad x+y=\Delta=\delta L_1,\qquad x=z\delta L_1,\qquad \mu\to0,\qquad \nu=\mu L_1\text{ fixed.}\tag{84}$$

*b. Explicit functional relations.* Using the nonzero-$\mu$ transfer matrices (83) we have, similarly to (58)

$$\begin{aligned}\hat{q}_+^{(2)}(x+L_1,y,s|L_1)&\propto\frac{\phi\tanh\mu}{\sinh(L_1\mu)}\frac{\phi\tanh\mu}{\sinh((L_1+y)\mu)}\\&\times\prod_{j=1}^{y-1}(1+\phi\tanh\mu\,\coth((L_1+j)\mu))\\&\times\prod_{j=1}^{x-1}\left(1+\frac{\phi\tanh\mu}{\sinh((L_1+y+j)\mu)}\gamma(j,y,\mu)^{-1}+\phi\tanh\mu\,\coth((L_1+y+j)\mu)\right),\\ \hat{q}_-^{(2)}(x+L_1,y,s|L_1)&\propto\frac{\phi\tanh\mu}{\sinh((L_1+x)\mu)}\\&\times\prod_{j=0}^{x-1}(1+\phi\tanh\mu\,\coth((L_1+j)\mu))\\&\times\prod_{j=1}^{y-1}\left(1+\frac{\phi\tanh\mu}{\sinh((L_1+x+j)\mu)}\gamma(x,j,\mu)+\phi\tanh\mu\,\coth((L_1+x+j)\mu)\right),\end{aligned}\tag{85}$$

where again $\gamma = \frac{\hat{q}_+^{(2)}}{\hat{q}_-^{(2)}}$.

### F. Scaling evaluation

Introducing $z, \delta, \nu = L_1\mu$ and replacing sums by integrals, Eq. (85) yields

$$\hat{q}_+^{(2)}(x+L_1,y,s|L_1) = \phi^2\mu^2 \frac{\sinh^{\phi-1}\nu}{\sinh^{\phi}((1+\delta)\nu)} \frac{\exp\big(\phi\nu\delta A_+(z)\big)}{\sinh((1+\delta(1-z))\nu)}, \tag{86}$$

with

$$A_+(z) = \int_0^z \frac{g\Big(\frac{u}{u+1-z}, \delta(1-z+u)\Big)^{-1}}{\sinh((1+(1-z+u)\delta)\nu)}\, du.$$

Similarly,

$$\hat{q}_-^{(2)}(x+L_1,y,s|L_1) = \phi\mu \frac{\sinh^{\phi}\nu}{\sinh^{\phi}((1+\delta)\nu)} \frac{\exp\big(\phi\nu\delta A_-(z)\big)}{\sinh((1+\delta z)\nu)}, \tag{87}$$

where

$$A_-(z) = \int_0^{1-z} \frac{g\Big(\frac{z}{z+u}, \delta(z+u)\Big)}{\sinh((1+(z+u)\delta)\nu)}\, du.$$

*a. Ratio equation.* The scaling function therefore satisfies

$$\begin{aligned} g = \frac{q_+}{q_-} = \frac{\phi\mu}{\sinh\nu} \frac{\sinh((1+\delta z)\nu)}{\sinh((1+\delta(1-z))\nu)} \\ \times \exp\Big[\phi\nu\delta\big(A_+(z) - A_-(z)\big)\Big]. \end{aligned} \tag{88}$$

*b. Final form.* Finally, it can be checked self-consistently that the following expression satisfies (86), (87) and recovers the integrated solution when $\nu \to 0$:

$$\boxed{g(z,\delta,\nu) = \frac{\phi\sinh(\delta z\nu)\,\sinh((1+\delta z)\nu)\,{}_2F_1\Big(1-\phi,\phi+1;2;\frac{\sinh((z-1)\delta\nu)\sinh(z\delta\nu)}{\sinh\nu\,\sinh((1+\delta)\nu)}\Big)}{\sinh\nu\,\sinh((1+\delta)\nu)\,{}_2F_1\Big(-\phi,\phi;1;\frac{\sinh((z-1)\delta\nu)\sinh(z\delta\nu)}{\sinh\nu\,\sinh((1+\delta)\nu)}\Big)}}. \tag{89}$$

It is surprising to see that it is simply the result (69) for $\mu = 0$, just "putting hyperbolic sines everywhere": this peculiar situation was already encountered in [4]. From (89), we finally recover the aged Ray–Knight results (19), (22).

## V. SOME CLOSED-FORM SIMPLE EXPRESSIONS FOR $\psi(x)$

In the main text, we discussed that for $\alpha = 2k+1$ an odd integer, the $\psi$ function can be expressed in terms of known functions. We give some examples here; these examples are easily computed using Mathematica, see the notebook on

the following GitHub page.

$$\psi_{3,2}(x) = \frac{2\sqrt{\frac{2}{\pi}}e^{-2/x}}{x^{5/2}}\left[e^{2/x}\Big(-16\sqrt{2\pi}\sqrt{x}\,\mathrm{erfc}\Big(\tfrac{\sqrt{2}}{\sqrt{x}}\Big) - 12\,\mathrm{Ei}\big(-\tfrac{2}{x}\big) + x^2\Big) - (x-10)x\right], \tag{90}$$

$$\psi_{3,3}(x) = \frac{3\sqrt{\frac{2}{\pi}}e^{-2/x}}{x^{3}}\left[e^{2/x}\Big(8\sqrt{2\pi}(4-5x)\,\mathrm{erfc}\Big(\tfrac{\sqrt{2}}{\sqrt{x}}\Big) + \sqrt{x}\Big(x^2 - 60\,\mathrm{Ei}\big(-\tfrac{2}{x}\big)\Big)\Big) - (x-16)(x-2)\sqrt{x}\right], \tag{91}$$

$$\psi_{5,2}(x) = \frac{2\sqrt{\frac{2}{\pi}}e^{-8/x}}{3x^{7/2}}\left[e^{8/x}\Big(128\sqrt{2\pi}\sqrt{x}\Big(\mathrm{erfc}\Big(\tfrac{\sqrt{2}}{\sqrt{x}}\Big) - 8\,\mathrm{erfc}\Big(\tfrac{2\sqrt{2}}{\sqrt{x}}\Big)\Big)\right. \tag{92}$$

$$+ 80\Big(\mathrm{Ei}\big(-\tfrac{2}{x}\big) - 16\,\mathrm{Ei}\big(-\tfrac{8}{x}\big)\Big) + 3x^3\Big) \tag{93}$$

$$\left. + x((x-12)x+352) - 4e^{6/x}x((x-3)x+22)\right], \tag{94}$$

$$\psi_{5,3}(x) = \frac{1}{\sqrt{\pi}x^4}\left\{16\left[35\sqrt{2}\sqrt{x}\Big(\mathrm{Chi}\big(\tfrac{2}{x}\big) - 16\,\mathrm{Chi}\big(\tfrac{8}{x}\big) - \mathrm{Shi}\big(\tfrac{2}{x}\big) + 16\,\mathrm{Shi}\big(\tfrac{8}{x}\big)\Big)\right.\right. \tag{95}$$

$$\left. + 8\sqrt{\pi}\Big((7x-4)\,\mathrm{erfc}\Big(\tfrac{\sqrt{2}}{\sqrt{x}}\Big) + 8(16-7x)\,\mathrm{erfc}\Big(\tfrac{2\sqrt{2}}{\sqrt{x}}\Big)\Big)\right] \tag{96}$$

$$+ \sqrt{2}e^{-8/x}\sqrt{x}\Big(3e^{8/x}x^3 + ((x-20)x+928)x \tag{97}$$

$$\left. - 4e^{6/x}(x((x-5)x+58)-64) - 4096\Big)\right\}. \tag{98}$$

In the above, Ei, Chi, Shi are respectively the exponential, cosine and sine integral functions [6].



\end{document}